\documentclass[aps,pra,amsmath,superscriptaddress,amssymb,reprint]{revtex4-1}

\usepackage{hyperref}
\hypersetup{
    colorlinks, linkcolor={blue},
    citecolor={blue}, urlcolor={blue},breaklinks
}
\usepackage{natbib}
\hypersetup{
    colorlinks, linkcolor={blue},
    citecolor={blue}, urlcolor={blue},breaklinks
}\usepackage{braket}
\usepackage{amsmath}
\usepackage{comment}
\usepackage{graphicx}
\usepackage{mathtools}
\usepackage{subcaption}

\newcommand*{\citen}[1]{%
  \begingroup
    \romannumeral-`\x 
    \setcitestyle{numbers}%
    \cite{#1}%
  \endgroup   
}

\newcommand{\nvec}[0]{\mathbf{n}}

\newcommand{\Pvec}[0]{\mathbf{P}}
\newcommand{\Qvec}[0]{\mathbf{Q}}
\newcommand{\zvec}[0]{\mathbf{z}}
\newcommand{\deriv}[2]{\frac{\dint #1}{\dint #2}}
\newcommand{\pderiv}[2]{\frac{\partial #1}{\partial #2}}
\newcommand{\Hord}[0]{H_{\text{ord}}}
\newcommand{\dint}[0]{\, \mathrm{d}}
\newcommand{\euler}[0]{\mathrm{e}}

\renewcommand{\thesubfigure}{(\roman{subfigure})}

\begin{document}

\title{Simulation of the Quantum Dynamics of Indistinguishable Bosons with the Coupled Coherent States Method}
\author{James A. Green}
\email{j.a.green1@leeds.ac.uk}
\affiliation{School of Chemistry, University of Leeds, Leeds LS2 9JT, United Kingdom}
\affiliation{Consiglio Nazionale delle Ricerche, Istituto di Biostrutture e Bioimmagini (CNR-IBB), via Mezzocannone 16, 80134, Napoli, Italy (Current Address)}

\author{Dmitrii V. Shalashilin}
\email{d.shalashilin@leeds.ac.uk}
\affiliation{School of Chemistry, University of Leeds, Leeds LS2 9JT, United Kingdom}

\date{\today}

\begin{abstract}
Computer simulations of many-body quantum dynamics of indistinguishable particles is a challenging task for computational physics. In this paper we demonstrate that the method of coupled coherent states (CCS) developed previously for multidimensional quantum dynamics of distinguishable particles can be used to study indistinguishable bosons in the second quantisation formalism. To prove its validity, the technique termed here coupled coherent states for indistinguishable bosons (CCSB) is tested on two model problems. The first is a system-bath problem consisting of a tunnelling mode coupled to a harmonic bath, previously studied by CCS and other methods in distinguishable representation in 20 dimensions. The harmonic bath is comprised of identical oscillators, and may be second quantised for use with CCSB, so that this problem may be thought of as a bosonic bath with an impurity. The cross-correlation function for the dynamics of the system and Fourier transform spectrum compare extremely well with a benchmark calculation, which none of the prior methods of studying the problem achieved. The second model problem involves 100 bosons in a shifted harmonic trap. Breathing oscillations in the 1-body density are calculated and shown to compare favourably to a multiconfigurational time-dependent Hartree for bosons calculation, demonstrating the applicability of the method as a new formally exact way to study the quantum dynamics of Bose-Einstein condensates.

\end{abstract}

\maketitle

\section{Introduction} 

In the past two decades there has been significant interest in systems of indistinguishable bosons, due to experimentally produced Bose-Einstein condensates of ultracold alkali metal atoms~\cite{anderson1995,bradley1995,davis1995}. These condensates, first posited by the eponymous Bose and Einstein in 1924-25, have permitted macroscopic observations of quantum phenomena and lead to a wealth of experimental research in areas such as atomic interferometry~\cite{orzel2001}, bosonic Josephson junctions~\cite{albiez2005,levy2007}, quantum vortices~\cite{matthews1999,madison2000} and the generation of solitons~\cite{burger1999,denschlag2000}.

From the theoretician's point of view, the Gross-Pitaevskii equation (GPE)~\cite{gross1961,pitaevskii1961} has been the predominant method used to study Bose-Einstein condensates, see for example Refs.~\cite{smerzi1997,perez-garcia1997,raghavan1999,bao2003,liang2005,ananikian2006} and the review articles~\cite{leggett2001,minguzzi2004}. However the GPE is a mean-field theory and as such cannot describe many-body effects in condensates. It also assumes that all bosons occupy a single state at all times, which is not the case during fragmentation. In recent years, the multiconfigurational time-dependent Hartree method for bosons (MCTDHB)~\cite{streltsov2007,alon2008} has been used to treat indistinguishable bosons from the standpoint of exact quantum mechanics~\cite{streltsov2008,streltsov2009,streltsov2009a,lode2009,sakmann2009,sakmann2010,streltsov2011,streltsov2011a,lode2012,streltsov2013,beinke2015}. A multi-layer version of MCTDHB has also been developed (ML-MCTDHB)~\cite{cao2013,kronke2013} that exploits the multi-layer structure to study mixed bosonic systems (for example impurities in Bose-Einstein condensates~\cite{schurer2017,keiler2018,mistakidis2018b,mistakidis2018c,katsimiga2018}, binary mixtures of Bose-Einstein condensates~\cite{mistakidis2018}, and solitons~\cite{katsimiga2017,kronke2015,katsimiga2018,katsimiga2017a}) and bosonic systems where different degrees of freedom may be separated (for example different spatial locations when bosons are residing in optical lattices~\cite{mistakidis2014,mistakidis2015,mistakidis2015a,mistakidis2017b,koutentakis2017,neuhaus-steinmetz2017,mistakidis2018a,plassmann2018}).

Before being used to treat indistinguishable bosons, standard MCTDH~\cite{meyer1990} and ML-MCTDH~\cite{wang2003,manthe2008} have been well established theories for treating distinguishable particles. They are able to solve the time-dependent Schr\"{o}dinger equation (TDSE) exactly for multiple degrees of freedom, albeit with basis sets that grow exponentially with increased dimensionality. Our own coupled coherent states (CCS) method has also demonstrated its propensity at solving the TDSE for distinguishable particles, with basis sets that scale more favourably with dimensionality~\cite{shalashilin2000,shalashilin2004}. This is achieved by using randomly sampled trajectory guided coherent states as basis functions, although the trade-off for this favourable scaling is that random noise and slow convergence may be present. Noise can cause a decay in auto/cross-correlation functions and may be reduced by increasing the number of configurations, or applying a filter diagonalisation technique to extract frequencies~\cite{shalashilin2003}. Importance sampling of coherent state basis set initial conditions is also key to the accuracy and efficiency of the CCS approach~\cite{shalashilin2008}.

In this present work we extend the CCS method to looking at indistinguishable bosons in the second quantisation representation, and dub the method coupled coherent states for indistinguishable bosons (CCSB). Due to the use of coherent states in CCSB and their relation to the creation and annihilation operators of second quantisation, together with the fact that systems with a large number of particles tend towards classical behaviour and the basis in CCSB is guided by classical-like trajectories, suggest that the method will be particularly suited to such systems. Indeed, recent semiclassical coherent state work with the Herman-Kluk method on indistinguishable bosons demonstrates this hypothesis~\cite{simon2014,ray2016}. CCSB is fully quantum however, as with standard CCS, and it has previously been shown that a local quadratic approximation of the Hamiltonian into the CCS equations yields the coherent state matrix of the Herman-Kluk propagator~\cite{child2003}. We anticipate that the CCSB method will provide a description of many-body dynamics over and above the mean-field Gross-Pitaevskii approach. Furthermore, as CCS has previously shown to be able to provide a similar numerical picture to MCTDH with lower computational scaling with dimensionality~\cite{shalashilin2004a}, we anticipate that CCSB may be able to do the same with respect to MCTDHB.

To illustrate the suitability of CCSB to problems involving indistinguishable bosons, we apply the method to two model problems. The first model problem consists of a bosonic bath with an impurity, demonstrating that the method is capable of studying multi-component bosonic systems and opening up the possibility of studying multi-atomic Bose-Einstein condensates~\cite{modugno2002}, spinor Bose-Einstein condensates~\cite{kawaguchi2012}, dark-bright solitons~\cite{becker2008}, and Bose-polarons~\cite{bruderer2007}. The second model problem consists of a collection of indistinguishable bosons in a harmonic trap, demonstrating the propensity of the method to study systems of bosons in optical lattices~\cite{greiner2002}, for example with the Bose-Hubbard model~\cite{jaksch2005,simon2014,ray2016}, and the possibility to study bosons in a single well that is deformed into a double well, such as that in Ref.~\cite{alon2008}, and observed in experimental bosonic Josephson junctions~\cite{gati2006,gati2007}.

\section{Numerical Details}

The CCSB method relies on the machinery of the CCS method, which has been derived and presented previously when treating distinguishable particles~\cite{shalashilin2000,shalashilin2004}. A description of the CCS method will be presented below, before a discussion on how the method is modified to treat indistinguishable bosons in the second quantisation representation in CCSB.

\subsection{Coupled Coherent States Working Equations}\label{sec:CCS_eqs}

In the CCS method, the wavefunction is represented as a basis set of trajectory guided coherent states, $\ket{z}$. The coordinate representation of a coherent state is given by
\begin{equation}
\braket{x | z} = \left(\frac{\gamma}{\pi}\right)^{1/4} \exp\left(-\frac{\gamma}{2}(x-q)^2 + \frac{i}{\hbar}p(x-q) + \frac{i p q}{2\hbar} \right),
\end{equation}
where $q$ and $p$ are the position and momentum centres of the coherent state, $\gamma$ is the width parameter of the coherent state, given by $\gamma = m \omega / \hbar$, with $m$ mass and $\omega$ frequency. In atomic units (which are used throughout the paper) $m = \omega = \hbar = 1$, thus $\gamma = 1$. Coherent states are eigenstates of the creation and annihilation operators respectively
\begin{subequations}
\label{eq:creat_ann_eig}
\begin{align}
\label{eq:creat_eig}
\bra{z} \hat{a}^{\dagger} &= \bra{z} z^* \\
\hat{a} \ket{z} &= z \ket{z},
\label{eq:ann_eig}
\end{align}
\end{subequations}
where the creation and annihilation operators are given by
\begin{subequations}
\label{eq:creat_ann}
\begin{align}
\label{eq:creat}
\hat{a}^{\dagger} &= \frac{1}{\sqrt{2}} \left( \hat{q} - i \hat{p} \right) \\
\hat{a} &= \frac{1}{\sqrt{2}} \left( \hat{q} + i \hat{p} \right).
\label{eq:ann}
\end{align}
\end{subequations}
The eigenvalues of Eqs.~\ref{eq:creat_eig} and~\ref{eq:ann_eig}, $z^*$ and $z$, can be used to label a coherent state, and from Eqs.~\ref{eq:creat} and~\ref{eq:ann} it can be seen they are given by
\begin{subequations}
\begin{align}
z^* &= \frac{1}{\sqrt{2}} \left( q - i p \right) \\
z &= \frac{1}{\sqrt{2}} \left( q + i p \right).
\end{align}
\end{subequations}
An important consequence of the above is that one may write a Hamiltonian in terms of creation and annihilation operators rather than position and momentum operators. A normal ordered Hamiltonian may then be obtained when the creation operators precede the annihilation ones
\begin{equation}
\hat{H}(\hat{q},\hat{p}) = \hat{H}(\hat{a},\hat{a}^\dagger) = \Hord(\hat{a}^\dagger,\hat{a}).
\end{equation}
From this, matrix elements of the Hamiltonian are simple to calculate in a coherent state basis
\begin{equation}\label{eq:H_CS}
\braket{z'|\Hord(\hat{a}^\dagger,\hat{a})|z} = \braket{z'|z}\Hord(z'^*,z),
\end{equation}
where the overlap $\braket{z'|z}$ is given by
\begin{equation}
\label{eq:CCS_ovrlp}
\braket{z'|z} = \exp\left(z'^* z - \frac{z'^* z'}{2} - \frac{z^* z}{2}\right).
\end{equation}
The wavefunction ansatz in CCS is given by
\begin{equation}
\label{eq:CCS_ansatz}
\ket{\Psi(t)} = \sum_{k=1}^{K} D_k(t) \euler^{i S_k(t)} \ket{z_k(t)},
\end{equation}
where the sum is over $K$ configurations, $D_k$ is a time dependent amplitude and $S_k$ is the classical action. The classical action in coherent state notation is given by
\begin{equation}
\label{eq:CCS_action}
S_k = \int \left[ \frac{i}{2} \left(z_k^* \dot{z}_k - \dot{z}_k^* z_k \right) - \Hord(z_k^*,z_k) \right] \dint t.
\end{equation}
The wavefunction is propagated via the time-dependence of the coherent state basis vectors, amplitudes and action. The coherent states are guided by classical trajectories, and evolve according to Hamilton's equation
\begin{equation}
\label{eq:CCS_zdot}
\dot{z}_k = -i \pderiv{\Hord(z^*_k,z_k)}{z^*_k}.
\end{equation}
The time-dependence of the amplitudes may be found via substitution of Eq.~\ref{eq:CCS_ansatz} into the time-dependent Schr\"{o}dinger equation and closing with a coherent state basis bra:
\begin{equation}
\label{eq:CCS_Ddot}
\sum_{l=1}^{K} \braket{z_k|z_l} \euler^{i S_l} \deriv{D_l}{t} = - i \sum_{l=1}^{K} \braket{z_k|z_l} \euler^{i S_l} D_l \delta^2 \Hord'(z_k^*,z_l),
\end{equation}
where the $\delta^2 \Hord'(z_k^*,z_l)$ term is
\begin{equation}
\label{eq:delta2_H}
\delta^2 \Hord'(z_k^*,z_l) = \Hord(z_k^*,z_l) - \Hord(z_l^*,z_l) - i \dot{z}_l(z_k^*-z_l^*).
\end{equation}
Finally, the time-dependence of the classical action is straightforwardly calculated from Eq.~\ref{eq:CCS_action}.

\subsection{Second Quantisation and CCSB}

CCS works for Hamiltonians that can be expressed via creation and annihilation operators in the normal ordered form as illustrated in Eq.~\ref{eq:H_CS}. In second quantisation, the Hamiltonian of a system of bosons also appears in a normal ordered form, therefore no modifications of the CCS working equations are required for treating indistinguishable bosons with CCSB. The only difference is that the coherent state basis functions are used to represent particle number occupations of quantum states in the second quantisation formalism, as opposed to individual particles in the distinguishable first quantisation representation.

In the second quantisation representation, multiparticle states are described in terms of an occupation number $n^{(\alpha)}$ that describes the number of particles belonging to a particular quantum state $\ket{\alpha}$. A Fock state describes the set of occupation number states
\begin{equation}
\ket{\nvec} = \prod_{\alpha=0}^{\Omega} \ket{n^{(\alpha)}} = \ket{n^{(0)},n^{(1)},\dots,n^{(\Omega)}},
\end{equation}
and may be generated by successive application of creation operators on the vacuum state~$\ket{0}$
\begin{equation}
\begin{split}
& \ket{n^{(0)}, n^{(1)}, \dots, n^{(\Omega)}} =
\\& \frac{\left(\hat{a}^{(0)\dagger}\right)^{n^{(0)}}}{\sqrt{n^{(0)}!}} \frac{\left(\hat{a}^{(1)\dagger}\right)^{n^{(1)}}}{\sqrt{n^{(1)}!}} \dots \frac{\left(\hat{a}^{(\Omega)\dagger}\right)^{n^{(\Omega)}}}{\sqrt{n^{(\Omega)}!}} \ket{0^{(0)},0^{(1)},\dots,0^{(\Omega)}}.
\end{split}
\end{equation}
In CCSB, the multidimensional version of the CCS wavefunction representation is used as a basis set expansion for Fock states
\begin{equation}
\ket{\nvec} = \sum_{k=1}^{K} D_k(t) \euler^{i S_k(t)} \ket{\zvec_k(t)},
\end{equation}
which is exactly analogous to Eq.~\ref{eq:CCS_ansatz}. The only difference is the multidimensional coherent state $\ket{\zvec_k}$  is a product of coherent states that describe occupations of each quantum state $\ket{\alpha}$
\begin{equation}
\ket{\zvec_k} = \prod_{\alpha=0}^{\Omega} \ket{z^{(\alpha)}}.
\end{equation}
Therefore any wavefunction in the basis of Fock states can be equivalently represented in the basis of coherent states. The Hamiltonian of a system of indistinguishable bosons can be second quantised and presented in terms of 1-body $\hat{h}(\Qvec)$, 2-body $\hat{W}(\Qvec,\Qvec')$, and creation and annihilation operators as
\begin{equation}\label{eq:2q_gen_H}
\begin{split}
\hat{H} = & \sum_{\alpha,\beta} \braket{\alpha|\hat{h}|\beta} \hat{a}^{(\alpha)\dagger} \hat{a}^{(\beta)} \\
& + \frac{1}{2} \sum_{\alpha,\beta,\gamma,\zeta} \braket{\alpha,\beta|\hat{W}|\gamma,\zeta} \hat{a}^{(\alpha)\dagger} \hat{a}^{(\beta)\dagger} \hat{a}^{(\zeta)} \hat{a}^{(\gamma)},
\end{split}
\end{equation}
where $\ket{\alpha}$, $\ket{\beta}$, $\ket{\gamma}$, and $\ket{\zeta}$ are quantum states. This conveniently gives a second quantised Hamiltonian in normal ordered form, which is required by CCSB. In the following sections CCSB is applied to two model problems to illustrate its ability to study fully quantum bosonic problems and compare to numerically exact results.

\section{Application 1: Double Well Tunnelling Problem}

The first application of CCSB is to an $M$-dimensional model Hamiltonian that consists of an $(M-1)$-dimensional harmonic bath, coupled to a 1-dimensional tunnelling mode governed by an asymmetric double well potential. This a system-bath problem, which may also be thought of as a bosonic bath with an impurity, previously studied in distinguishable representation with linear coupling of the bath to the system by matching pursuit split-operator Fourier transform (MP/SOFT)~\cite{wu2004}, standard CCS~\cite{sherratt2006}, a trajectory guided configuration interaction (CI) expansion of the wavefunction~\cite{habershon2012}, an adaptive trajectory guided (aTG) scheme~\cite{saller2017}, Gaussian process regression (GPR)~\cite{alborzpour2016}, and a basis expansion leaping multi-configuration Gaussian (BEL MCG) method~\cite{murakami2018}. It has also been studied with quadratic coupling of the bath to the system by MP/SOFT~\cite{wu2004}, standard CCS~\cite{sherratt2006}, trajectory guided CI~\cite{habershon2012}, aTG~\cite{saller2017} and a 2-layer version of CCS (2L-CCS)~\cite{green2016}. A benchmark calculation for the quadratic coupling case has also been proposed in recent work~\cite{green2015}, using a relatively simple wavefunction expansion in terms of particle in a box wavefunctions for the tunnelling mode, and harmonic oscillator wavefunctions for the harmonic bath. The size of the calculation in Ref.~\cite{green2015} was greatly reduced by exploiting the indistinguishability of the bath configurations, the first time this had been considered, and a well converged result was achieved, prompting the idea of CCSB. The quadratic coupling case is the one we consider in this application.

The Hamiltonian is given in distinguishable representation by
\begin{equation}\label{eq:H_AS}
    \hat{H}
    = \frac{\hat{p}^{(1)^2}}{2} - \frac{\hat{q}^{(1)^2}}{2} + \frac{\hat{q}^{(1)^4}}{16 \eta}
    + \frac{\hat{\mathbf{P}}^2}{2} + \frac{\left(1+\lambda \hat{q}^{(1)}\right) \hat{\mathbf{Q}}^2}{2},
\end{equation}
where $(\hat{q}^{(1)},\hat{p}^{(1)})$ are the position and momentum operators of the 1-dimensional system tunnelling mode, and $(\hat{\mathbf{Q}},\hat{\mathbf{P}})$ are the position and momentum operators of the $(M-1)$-dimensional harmonic bath modes, with $\hat{\mathbf{Q}}=\sum_{m=2}^{M} \hat{q}^{(m)}$ and $\hat{\mathbf{P}}=\sum_{m=2}^{M} \hat{p}^{(m)}$. The coupling between system and bath is given by the constant $\lambda$, whilst $\eta$ determines the well depth.

In previous work~\cite{wu2004,sherratt2006,habershon2012,green2015,green2016,saller2017}, the parameters $\lambda=0.1$ and $\eta=1.3544$ have been used in a 20-dimensional ($M=20$) problem, which we also consider. The initial wavefunction $\ket{\Psi(0)}$ is a multidimensional Gaussian wavepacket, with initial position and momentum centres for the tunnelling mode $\hat{q}^{(1)}(0) = -2.5$ and $\hat{p}^{(1)}(0) = 0.0$, and for the bath modes $\hat{q}^{(m)}(0) = 0.0$ and $\hat{p}^{(m)}(0) = 0.0$ $\forall$ $m$.

As the bath oscillators have the same initial conditions and the same frequency, they can be thought of as indistinguishable, and the bath part of the Hamiltonian may be second quantised for use with CCSB. As the tunnelling mode is not part of this indistinguishable system, the portion of the Hamiltonian that describes it will not be second quantised. However, this will not pose a problem as the dynamical equations are identical for CCS and CCSB, the only subtlety is the interpretation of the coherent state basis vectors $\ket{\zvec}$ as will be discussed below. Using the definition of a second quantised Hamiltonian in Eq.~\ref{eq:2q_gen_H}, and the definition of coherent states as eigenstates of the creation and annihilation operators, Eq.~\ref{eq:H_AS} may be written in normal-ordered form as $\Hord(\hat{a}^{\dagger},\hat{a})$, for which the coherent state matrix element $\braket{\zvec_k | \Hord(\hat{a}^{\dagger},\hat{a}) | \zvec_l} = \braket{\zvec_k | \zvec_l} \Hord(\zvec_{k}^*,\zvec_{l})$, where
\begin{widetext}
\begin{equation}
\begin{split}
\Hord(\zvec_{k}^*,\zvec_{l}) = & - \frac{1}{2} \left(z^{(m=1)*^2}_{k} + z^{(m=1)^2}_{l}\right) + \frac{1}{64 \eta} \left(z^{(m=1)*^4}_{k} + z^{(m=1)^4}_{l} + 4 z^{(m=1)*^3}_{k} z^{(m=1)}_{l} + 4 z^{(m=1)*}_{k} z^{(m=1)^3}_{l} \right. \\
& \left. + 6 z^{(m=1)*^2}_{k} z^{(m=1)^2}_{l} + 12 z^{(m=1)*}_{k} z^{(m=1)}_{l} + 6 z^{(m=1)*^2}_{k} + 6 z^{(m=1)^2}_{l} + 3 \right) \\
& + \sum_{\alpha=0}^{\Omega}  z^{(2\alpha)*}_k z^{(2\alpha)}_l \epsilon^{(2\alpha)} + \frac{\lambda}{2} \sum_{\alpha,\beta=0}^{\Omega} z^{(2\alpha)*}_k z^{(2\beta)}_l Q^{(2\alpha,2\beta)^2} \left(z^{(m=1)*}_{k} + z^{(m=1)}_{l} \right).
\end{split}
\label{eq:Hord_1}
\end{equation}
\end{widetext}
The quantum states $\ket{\alpha}$ and $\ket{\beta}$ in Eq.~\ref{eq:Hord_1} are those of the harmonic oscillator with $\alpha$ and $\beta$ numbers of quanta, $\epsilon^{(\alpha)}$ is the eigenvalue for $\ket{\alpha}$, and the position and momentum operators of the tunnelling mode have explicitly been labelled with $(m=1)$ to distinguish them from the $\alpha$ labelling scheme of the second quantised bath modes. A full derivation of this, alongside evaluation of the $Q^{(2\alpha,2\beta)^2}$ matrix element is shown in Appendix~\ref{sec:2Q_H1}. Note that only even harmonic oscillator levels are required due to all bath modes initially residing in the ground level, as previously assumed~\cite{green2015}, and the bath having quadratic coupling to the system meaning only even harmonic oscillator levels will be occupied.

The multidimensional coherent state basis vector $\ket{\zvec}$ is represented as
\begin{equation}
\ket{\zvec} = \ket{z^{(m=1)}} \times \prod_{\alpha=0}^{\Omega} \ket{z^{(2\alpha)}},
\end{equation}
where $\ket{z^{(m=1)}}$ is a basis function for the tunnelling mode and $\ket{z^{(2\alpha)}}$ is a basis function for the second quantised bath modes. The determination of initial conditions for these coherent state basis functions, as well as the values of the initial amplitudes is shown in the following section.

\subsection{Initial Conditions for Application 1}\label{sec:init1}

The initial coherent state basis functions for the tunnelling mode are sampled from a Gaussian distribution centered around the initial tunnelling mode coordinates and momenta, as in previous works~\cite{sherratt2006,green2016}
\begin{equation}\label{eq:mc_sampling}
f(z^{(m=1)}) \propto \exp \left(-\sigma^{(m=1)} \left|z^{(m=1)} - z^{(m=1)}(0)\right|^2\right),
\end{equation}
where $\sigma^{(m=1)}$ is a parameter governing the width of the distribution. 

Sampling the initial coherent states for the bath can be performed by obtaining a probability distribution from the square of the coherent state representation of the initial bath Fock state. The initial bath Fock state is equal to 
\begin{equation}
\begin{split}
\ket{\nvec} &= \prod_{\alpha=0}^{\Omega} \ket{n^{(2\alpha)}} \\
&= \ket{n^{(2\alpha=0)},n^{(2\alpha=2)},\dots,n^{(2\alpha=2\Omega)}} \\
&= \ket{(M-1),0,\dots,0},
\end{split}
\end{equation}
where there are $M-1$ bath oscillators all in the ground harmonic oscillator state. Using the representation of a coherent state in a basis of Fock states
\begin{equation}\label{eq:CS_Fock}
\ket{z} = \euler^{-\frac{|z|^2}{2}} \sum_{n^{(\alpha)}} \frac{z^{n^{(\alpha)}}}{\sqrt{(n^{(\alpha)}!)}} \ket{n^{(\alpha)}}
\end{equation}
the following may be obtained
\begin{equation}
|\braket{z^{(2\alpha)} | n^{(2\alpha)}}|^2 = \frac{ \euler^{-|z^{(2\alpha)}|^2} \left(|z^{(2\alpha)}|^2\right)^{n^{(2\alpha)}}}{\pi n^{(2\alpha)}!} ,
\end{equation}
where the value of $\pi$ has appeared to enforce normalisation. This resembles a Poissonian distribution, however $|z^{(2\alpha)}|^2$ is continuous so a gamma distribution is used instead
\begin{equation}\label{eq:gam_dist}
f(|z^{(2\alpha)}|^2) \propto \frac{\left(|z^{(2\alpha)}|^2\right)^{n^{(2\alpha)}} \euler^{\frac{-|z^{(2\alpha)}|^2}{\sigma^{(2\alpha)}}}}{\Gamma(n^{(2\alpha)} + 1) \left(\sigma^{(2\alpha)}\right)^{n^{(2\alpha)} + 1}} ,
\end{equation}
where $\sigma^{(2\alpha)}$ is a compression parameter controlling the width of the distribution, and $\Gamma$ is the gamma function that is calculated using $n^{(2\alpha)} + 1$ because $\Gamma(n) = (n-1)!$.

The gamma distribution will be centred around $\sigma^{(2\alpha)} n^{(2\alpha)}$, however $|\braket{z^{(2\alpha)} | n^{(2\alpha)}}|^2$ should be centred around $|z^{(2\alpha)}|^2 = n^{(2\alpha)}$ as its maximum is found by
\begin{widetext}
\begin{equation}
\deriv{|\braket{z^{(2\alpha)} | n^{(2\alpha)}}|^2}{|z^{(2\alpha)}|^2}  =  \frac{1}{\pi n^{(2\alpha)}!} \left(-\euler^{-|z^{(2\alpha)}|^2}  \left(|z^{(2\alpha)}|^2\right)^{n^{(2\alpha)}} + n^{(2\alpha)} \euler^{-|z^{(2\alpha)}|^2} \left(|z^{(2\alpha)}|^2\right)^{n^{(2\alpha)}-1} \right) = 0.
\end{equation}
\end{widetext}
Fortunately, this is not an issue, as when $n^{(2\alpha)}=0$ for states $2\alpha>0$, the distribution will be centred around 0 irrespective of the compression parameter, and for ${n^{(2\alpha=0)}=M-1}$ a compression parameter of ${\sigma^{(2\alpha=0)}=1.0}$ is used. As we are not constrained by a choice of compression parameter  ${\sigma^{(2\alpha>0)}}$ for the states $2\alpha>0$, we are free to alter it to influence the accuracy of the calculation, and the final result presented in the following section uses ${\sigma^{(2\alpha>0)}} = 100$. The affect of altering this parameter is discussed in Sec.~\ref{sec:app1_conv}.

The initial amplitudes are calculated by projection of the initial basis onto the initial wavefunction with the action set to zero
\begin{equation}\label{eq:init_bigD_2Q_CCSB}
\braket{\zvec_k(0) | \Psi(0)} = \sum_{l=1}^{K} D_l(0) \braket{\zvec_k(0)|\zvec_l(0)}.
\end{equation}
The overlap of the initial coherent state basis with the initial wavefunction can be decomposed to
\begin{equation}
\braket{\zvec_k(0) | \Psi(0)} = \braket{z_k^{(m=1)}(0) | \Psi^{(m=1)}(0) } \braket{\prod_{\alpha=0}^\Omega z_k^{(2\alpha)}(0) | \nvec)}.
\end{equation}
The coherent state overlap with initial tunnelling mode wavefunction $\braket{z_k^{(m=1)}(0) | \Psi^{(m=1)}(0) }$ can be calculated via a Gaussian overlap, Eq.~\ref{eq:CCS_ovrlp}, using the initial positions and momenta for the tunnelling mode $\hat{q}^{(m=1)}(0) = -2.5$ and $\hat{p}^{(m=1)}(0) = 0.0$. The coherent state overlap with initial bath Fock state can be calculated by once more using the coherent state representation in a basis of Fock states, Eq.~\ref{eq:CS_Fock}
\begin{equation}
\begin{split}
\braket{\prod_{\alpha=0}^\Omega z_k^{(2\alpha)}(0) | \nvec)} = \left[\prod_{\alpha=0}^\Omega \euler^{-\frac{|z_k^{(2\alpha)}(0)|^2}{2}} \right] \frac{(z_k^{(2\alpha=0) *}(0))^{M-1}}{\sqrt{(M-1)!}}.
\end{split}
\label{eq:AS_2q_z_n_ovlp}
\end{equation}

\subsection{Results and Comparison to Other Methods}

\begin{figure*}
\centering
 \begin{subfigure}{0.48\textwidth}
   (i)

   \includegraphics[width=0.85\textwidth]{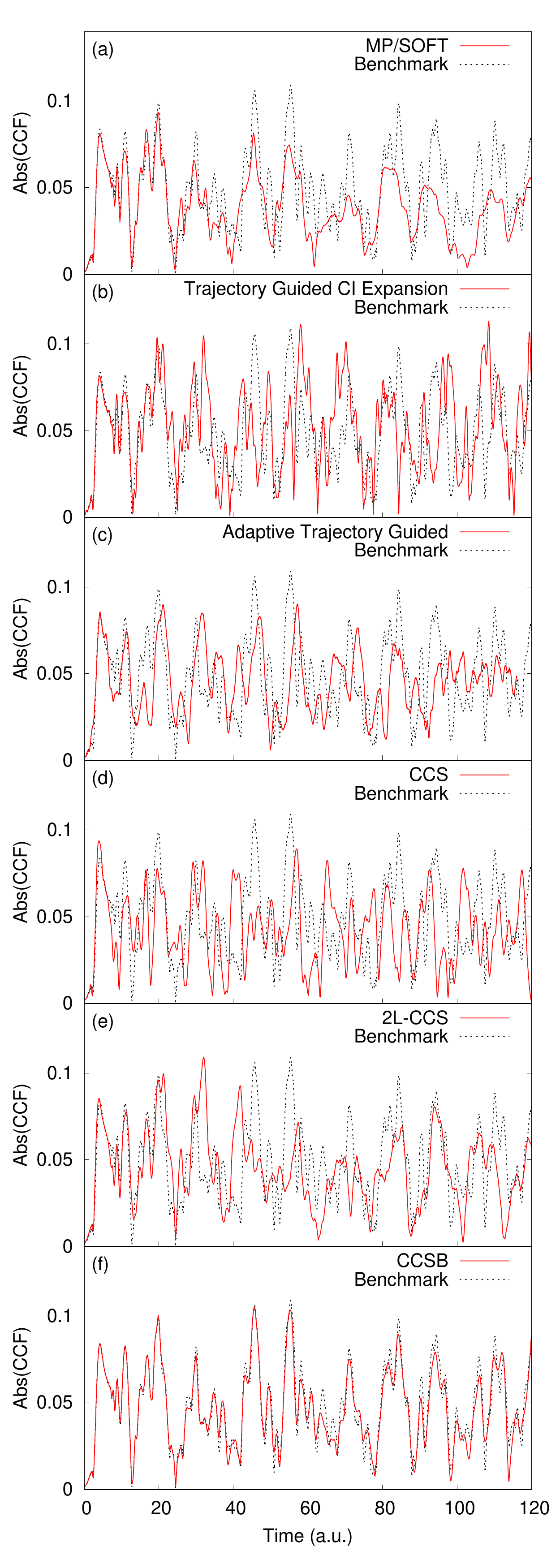}
   \phantomcaption
   \label{fig:ccf_20D}
  \end{subfigure}
 \begin{subfigure}{0.48\textwidth}
   (ii)

   \includegraphics[width=0.85\textwidth]{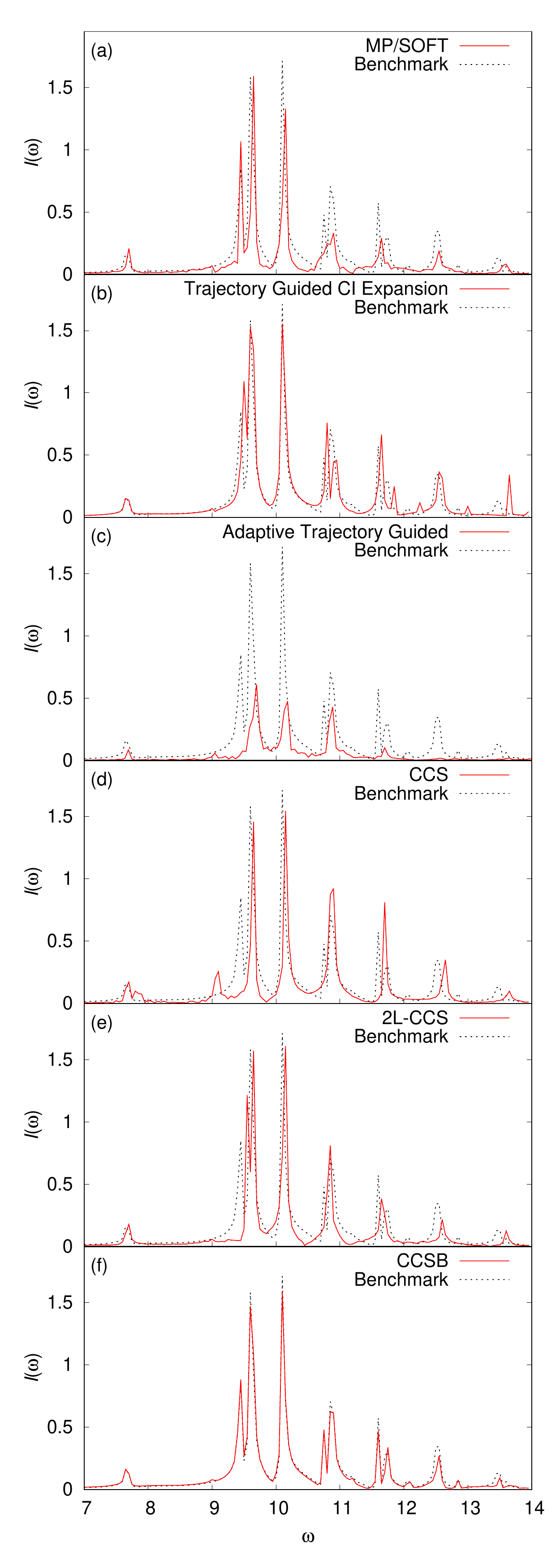}
   \phantomcaption
   \label{fig:ft_20D}
 \end{subfigure}
\caption{Comparison of (i) absolute values of the cross-correlation functions and (ii) Fourier transforms of the real part of the cross-correlation functions for different methods (red, solid) of studying Eq.~\ref{eq:H_AS} with $M=20$. $\lambda=0.1$ parameters relative to the benchmark~\cite{green2015} (black, dotted): (a) MP/SOFT~\cite{wu2004}, (b) Trajectory Guided CI Expansion~\cite{habershon2012}, (c) aTG~\cite{saller2017}, (d) CCS~\cite{sherratt2006}, (e) 2L-CCS~\cite{green2016}, (f) CCSB (present work).}
\label{fig:ccf_ft_20D}
\end{figure*}

The quantity of interest used to assess the performance of CCSB and compare it to previous methods of studying the problem~\cite{wu2004,sherratt2006,habershon2012,green2015,green2016,saller2017} is the cross-correlation function (CCF). This is the overlap between the wavefunction at time $t$ and the mirror image of the initial wavepacket, $\ket{\bar{\Psi}(0)}$, i.e. $\braket{\bar{\Psi}(0)|\Psi(t)}$. The mirror image of the initial state has coordinates for the tunnelling mode of $\bar{q}^{(1)}(0) = +2.5$ and $\bar{p}^{(1)}(0) = 0.0$, with bath modes in the ground harmonic level. It is located in the upper well of the asymmetric double well tunnelling potential, therefore non-zero values of the CCF are indicative of tunnelling. The spectrum of the CCF is also presented via a Fourier transform (FT) of the real part of the CCF.

The results of the CCSB calculation compared with previous methods of studying the 20D, $\lambda=0.1$ case~\cite{wu2004,sherratt2006,habershon2012,green2015,green2016,saller2017} is shown in Fig.~\ref{fig:ccf_ft_20D} with the absolute values of the CCFs in Fig.~\ref{fig:ccf_20D}, and FT spectra in Fig.~\ref{fig:ft_20D}. As can be seen from these two figures, the CCSB results compare extremely favourably to the benchmark calculation, with much closer agreement than prior methods. Previously the trajectory guided CI expansion was the closest result to the benchmark, due to its basis set expansion of time-independent basis functions used to represent excited state configurations being similar to the benchmark approach. However the CCF still differed from the benchmark, with noticeable differences occurring after 25 a.u., possibly due to approximations used in sampling the potential energy surface, despite the FT obtaining splitting of the higher energy peaks that no prior method managed. For this present CCSB calculation, there is no significant degradation of the calculation at $t>25$ a.u. as with the other methods, and the splitting of the high energy peaks is very well reproduced. As was alluded to in Ref.~\cite{green2016}, for this Hamiltonian a detailed description of the bath is required for accurate propagation, which is achieved in CCSB by taking account of the symmetry of the Hamiltonian.

The CCSB calculation uses $K=4000$ configurations and $\Omega=5$ even harmonic oscillator levels in the bath basis. The dimensionality of this problem has therefore been reduced from 20 to 6. The influence on the CCSB calculation of altering these parameters, as well as the compression parameter chosen of $\sigma^{(2\alpha>0)}=100$ is shown in the following section.

\subsection{Numerical Accuracy and Convergence}\label{sec:app1_conv}

\renewcommand{\thesubfigure}{(\roman{subfigure})}

\begin{figure*}
\centering
 \begin{subfigure}{0.48\textwidth}
   (i)

   \includegraphics[width=0.85\textwidth]{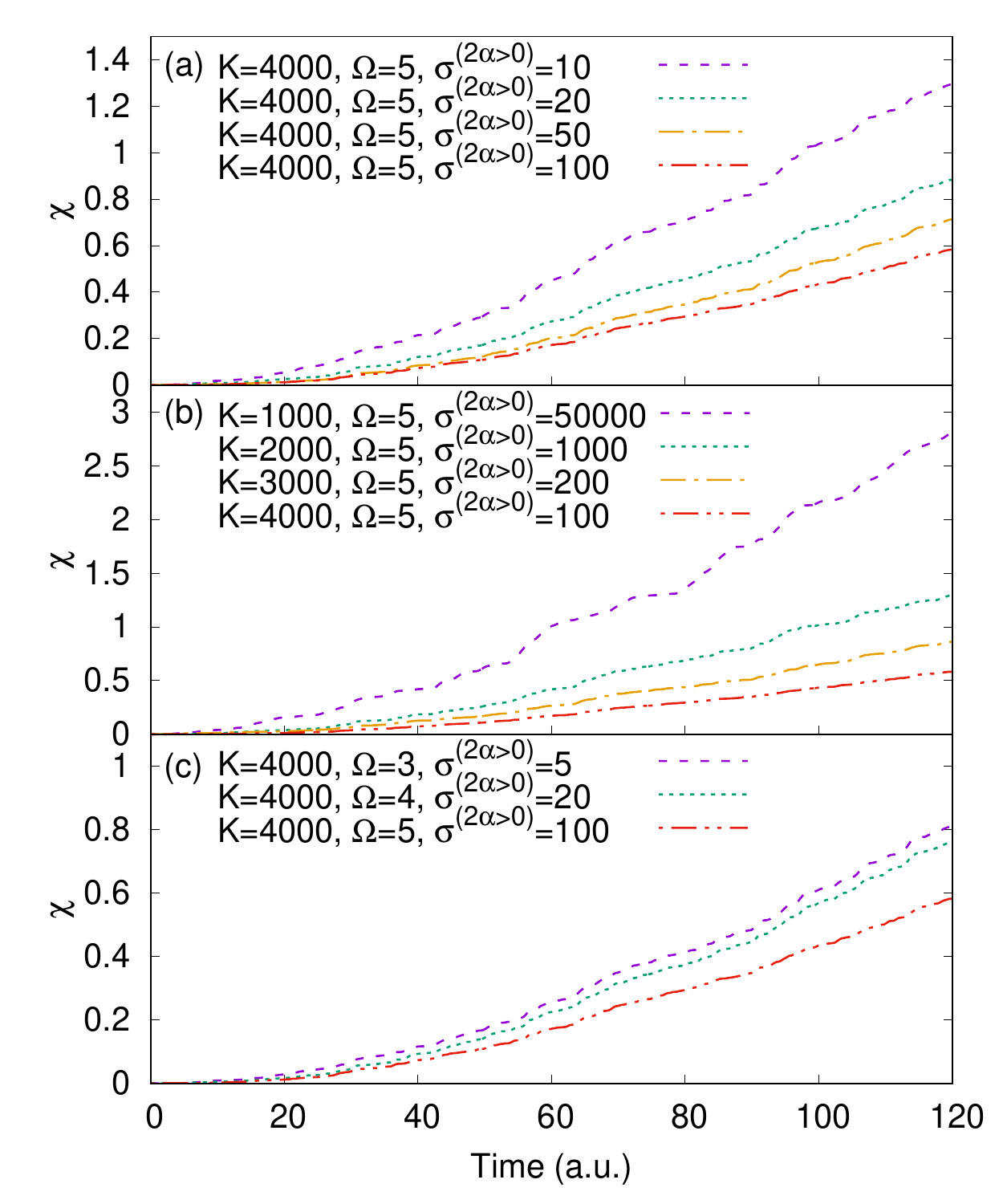}
   \phantomcaption
   \label{fig:abs_ccf_diff}
  \end{subfigure}
 \begin{subfigure}{0.48\textwidth}
   (ii)

   \includegraphics[width=0.85\textwidth]{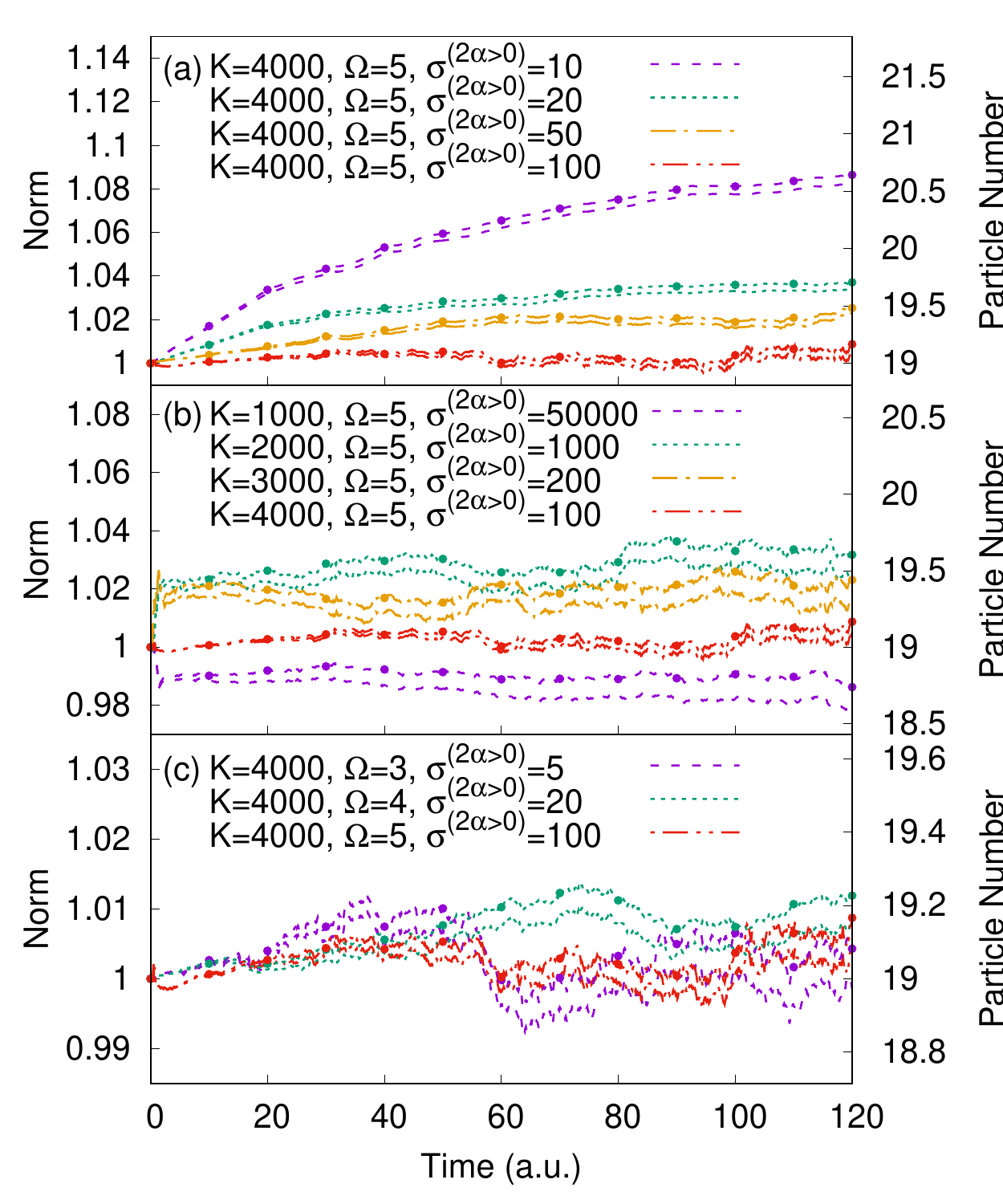}
   \phantomcaption
   \label{fig:norm_part_num}
 \end{subfigure}
\caption{(i) Cumulative error $\chi$ (defined in Eq.~\ref{eq:chi_err}) of the CCSB method with respect to the benchmark~\cite{green2015} for different values of: (a) compression parameter for coherent state sampling of the bath basis levels with zero initial occupation $\sigma^{(2\alpha>0)}$, (b) configurations $K$, and (c) even harmonic oscillator levels in bath basis $\Omega$ . (ii) Norm (dashed/dotted lines without circles) and particle number (dashed/dotted lines with circles) of CCSB calculations with different values of: (a) $\sigma^{(2\alpha>0)}$, (b) $K$, and (c) $\Omega$. Note that in panels (b) and (c) for both (i) and (ii) the value of  $\sigma^{(2\alpha>0)}$ changes as well as $K$ and $\Omega$. This is addressed in the text.}
\end{figure*}

Using an approach first presented in Ref.~\cite{green2016} to illustrate the accuracy and convergence of a method studying Application 1 with respect to the benchmark calculation, we define an error parameter $\chi$ as
\begin{equation}\label{eq:chi_err}
\begin{split}
\chi = \int & | \text{Abs}(\braket{\bar{\Psi}(0)|\Psi(t)})_{\text{bench}} \\
& - \text{Abs}(\braket{\bar{\Psi}(0)|\Psi(t)})_{\text{CCSB}} | \dint t,
\end{split}
\end{equation}
which indicates the cumulative error of the absolute value of the cross-correlation function of the CCSB method compared to the benchmark. This is shown in Fig.~\ref{fig:abs_ccf_diff} for different values of $\sigma^{(2\alpha>0)}$, $K$ and $\Omega$, in panels (a), (b) and (c).

As with CCS, the CCSB method does not conserve the norm $\braket{\Psi(t)|\Psi(t)}$ by default due to the use of a basis consisting of a superposition of coherent states~\cite{shalashilin2004}. However, this can be a useful property as the extent of norm conservation can be used to determine the accuracy and reliability of a propagation. Another important quantity for CCSB to conserve is the total particle number 
\begin{equation}
\begin{split}
N &= \braket{\Psi(t)|\sum_{\alpha=0}^{\Omega} \hat{a}^{(\alpha)\dagger} \hat{a}^{(\alpha)}|\Psi(t)} \\
& = \sum_{k,l=1}^K \sum_{\alpha=0}^{\Omega} D_k^* D_l \euler^{i(S_l - S_k)} \braket{\zvec_k|\zvec_l} z_k^{(\alpha)*} z_l^{(\alpha)},
\end{split}
\end{equation}
which for Application 1 amounts to the number of oscillators in the bath, $N=(M-1)=19$. Plots of the norm and particle number conservation for different values of  $\sigma^{(2\alpha>0)}$, $K$, and $\Omega$, (as in Fig.~\ref{fig:abs_ccf_diff}), are shown in Fig.~\ref{fig:norm_part_num}, with the value of the norm given by the dashed/dotted lines without circles and the particle number by the dashed/dotted lines with circles. It can be seen that the values of the norm and particle number follow each other closely for all calculations, and we will discuss the specific cases in the following.

Firstly, considering panel (a) in Figs.~\ref{fig:abs_ccf_diff} and~\ref{fig:norm_part_num}, both $K$ and $\Omega$ are held fixed whilst $\sigma^{(2\alpha>0)}$ is varied. It can be seen that the quality of the calculation with respect to the error term $\chi$, and the conservation of the norm and particle number improves with increasing $\sigma^{(2\alpha>0)}$. Further increase of $\sigma^{(2\alpha>0)}$ results in a numerically unstable propagation, as the value of the norm and particle number explodes as the basis is overcompressed. This suggests that appropriate choice of $\sigma^{(2\alpha>0)}$ is necessary for the initial sampling of the coherent states, as a value that is too small leads to errors due to the coherent states spreading too quickly, whilst a value that is too large leads to numerical instability.

Secondly, considering panel (b) in Figs.~\ref{fig:abs_ccf_diff} and~\ref{fig:norm_part_num}, the value of $K$ is varied whilst $\Omega$ is held constant. The value of $\sigma^{(2\alpha>0)}$ was chosen based on the criteria presented in the previous paragraph, with larger values of $\sigma^{(2\alpha>0)}$ for smaller values of $K$. This phenomenon has been noted in previous studies with CCS, where larger compression parameters are necessary for basis sets with fewer configurations, see Ref.~\cite{shalashilin2008} for further details. The error $\chi$ decreases with increasing $K$, as Monte-Carlo noise causing decay of the cross-correlation function decreases with increasing number of configurations. However, $\chi$ does not remain at 0 for the duration of the calculation, and the $K=4000$ propagation is only equivalent to the $K=3000$ calculation for the first 30 a.u. This indicates the slow convergence of the method as mentioned in the introduction. However, we can regard the accuracy of the $K=4000$ calculation as sufficient for this application as we are able to obtain an accurate FT spectrum with correct frequencies, alongside the good conservation of norm and particle number compared to the other calculations with fewer configurations.

Finally, considering panel (c) in  Figs.~\ref{fig:abs_ccf_diff} and~\ref{fig:norm_part_num}, the value of $\Omega$ is varied whilst $K$ is held constant. The value of $\sigma^{(2\alpha>0)}$ was again chosen based on the criteria presented above, with larger values possible with increased $\Omega$. It can be seen that altering the value of $\Omega$ has a small effect on the accuracy of the calculation (note the difference in y axis values for $\chi$ compared to panels (a) and (b)), and the value $\Omega=5$ was deemed to result in a stable enough propagation.

\section{Application 2: Indistinguishable Bosons in a Displaced Harmonic Trap}\label{sec:BEC}

The second application of CCSB is to a system composed purely of indistinguishable bosons, with $N$ interacting bosons placed in a harmonic trap displaced from the origin, with $N=100$ used in the present application. The oscillations in the density are calculated and compared to MCTDHB~\cite{streltsov2007,alon2008} calculations (performed by the authors, using the MCTDHB package~\cite{mctdhb_lab}). The Hamiltonian (in dimensionless units and distinguishable representation) for this problem consists of a shifted harmonic potential and a 2-body interaction term
\begin{equation}\label{eq:H_BC}
\hat{H} = \frac{\hat{\Pvec}^2}{2} + \frac{(\hat{\Qvec}-\xi)^2}{2} + \hat{W}(\Qvec,\Qvec'), 
\end{equation}
where $\hat{\Qvec}$ and $\hat{\Pvec}$ are the position and momentum operators of the $N$ bosons, $\xi=2.1$ is a parameter that shifts the harmonic potential from the origin, and $\hat{W}$ is the 2-body interaction, given by the contact interaction
\begin{equation}
\hat{W}(\Qvec,\Qvec') = \lambda_0 \delta(\Qvec-\Qvec').
\end{equation}
The constant $\lambda_0$ controls the strength of the interaction, with values of  $\lambda_0=0.001$ and  $\lambda_0=0.01$ used in the present application, whilst $\delta(\Qvec-\Qvec')$ is the Dirac delta function. As with Application 1, the Hamiltonian in Eq.~\ref{eq:H_BC} must be second quantised and normal-ordered before it can be used with CCSB, with
\begin{widetext}
\begin{equation}
\label{eq:Hord_2}
\Hord(\zvec_{k}^*,\zvec_{l}) =   \sum_{\alpha=0}^{\Omega} \epsilon^{(\alpha)} z_k^{(\alpha)*} z_l^{(\alpha)} - \sum_{\alpha,\beta=0}^{\Omega}  \xi Q^{(\alpha,\beta)} z_k^{(\alpha)*} z_l^{(\beta)} + \sum_{\alpha=0}^{\Omega} \frac{\xi^2}{2} z_k^{(\alpha)*} z_l^{(\alpha)} + \frac{1}{2} \sum_{\alpha,\beta,\gamma,\zeta=0}^{\Omega} \lambda_0 \delta^{(\alpha,\beta,\gamma,\zeta)} z_k^{(\alpha)*} z_k^{(\beta)*} z_l^{(\zeta)} z_l^{(\gamma)}.
\end{equation}
\end{widetext}
The derivation of the above, and evaluation of the matrix elements $Q^{(\alpha,\beta)}$ and $\delta^{(\alpha,\beta,\gamma,\zeta)}$ is shown in Appendix~\ref{sec:2Q_H2}. The initial sampling of the coherent states and amplitudes is performed in a similar manner to the second quantised bath of Application 1, and is shown in the following section.

\subsection{Initial Conditions for Application 2}\label{sec:init2}

The initial Fock state for the system includes all bosons in the ground harmonic state
\begin{equation}
\label{eq:init_n}
\ket{\nvec} = \prod_{\alpha=0}^{\Omega} \ket{n^{(\alpha)}} = \ket{n^{(0)},n^{(1)},\dots,n^{(\Omega)}} = \ket{100,0,\dots,0}.
\end{equation}

As with the second quantised bath of Application 1, the coherent states are sampled via a gamma distribution like in Eq.~\ref{eq:gam_dist}. The ground state with initial occupation $n^{(\alpha=0)}=100$ is sampled with compression parameter $\sigma^{(\alpha=0)}=1.0$ to ensure the distribution is centred in the correct place, whilst once more we are free to choose the compression parameter for the excited states with initial occupation $n^{(\alpha>0)}=0$. Values of $\sigma^{(\alpha>0)}=10^9$ for $\lambda_0=0.001$ and $\sigma^{(\alpha>0)}=10^7$ for $\lambda_0=0.01$ are used, with full details for the determination of these compression parameters shown in the following section.

Initial amplitudes are calculated by projecting the basis onto the initial Fock state in Eq.~\ref{eq:init_n}
\begin{equation}\label{eq:init_bigD_BEC_CCSB}
\braket{\zvec_k(0) |\nvec } = \sum_{l=1}^{K} D_l(0) \braket{\zvec_k(0)|\zvec_l(0)},
\end{equation}
where
\begin{equation}
\begin{split}
\braket{\zvec_k(0) |\nvec } &= \braket{\prod_{\alpha=0}^\Omega z_k^{(\alpha)}(0) | \nvec)} \\
& = \left[\prod_{\alpha=0}^\Omega \euler^{-\frac{|z_k^{(\alpha)}(0)|^2}{2}} \right] \frac{(z_k^{(\alpha=0) *}(0))^{100}}{\sqrt{100!}}.
\end{split}
\label{eq:BEC_z_n_ovlp}
\end{equation}

For the MCTDHB calculations, the initial orbitals were constructed from eigenfunctions of the unshifted trap ($\xi=0$), with the coefficient of one of the orbitals set to 1, whilst the rest were set to 0. This was chosen for the initial conditions of the MCTDHB calculations rather than propagation in imaginary time to obtain the initial orbitals and coefficients of the ground state~\cite{streltsov2007,alon2008}, as we currently do not have an analogous procedure for CCSB due to the instability of trajectories when propagating in imaginary time~\cite{shalashilin2004b}. This way we ensure the initial conditions for both methods are the same, and we are testing the propagation accuracy of both methods. In future work we will look at the effect of initial conditions on CCSB, and its comparison to MCTDHB.

\subsection{Results and Comparison to MCTDHB}

\begin{figure*}
\centering
\begin{subfigure}{0.48\textwidth}
(i)

\includegraphics[width=0.9\textwidth]{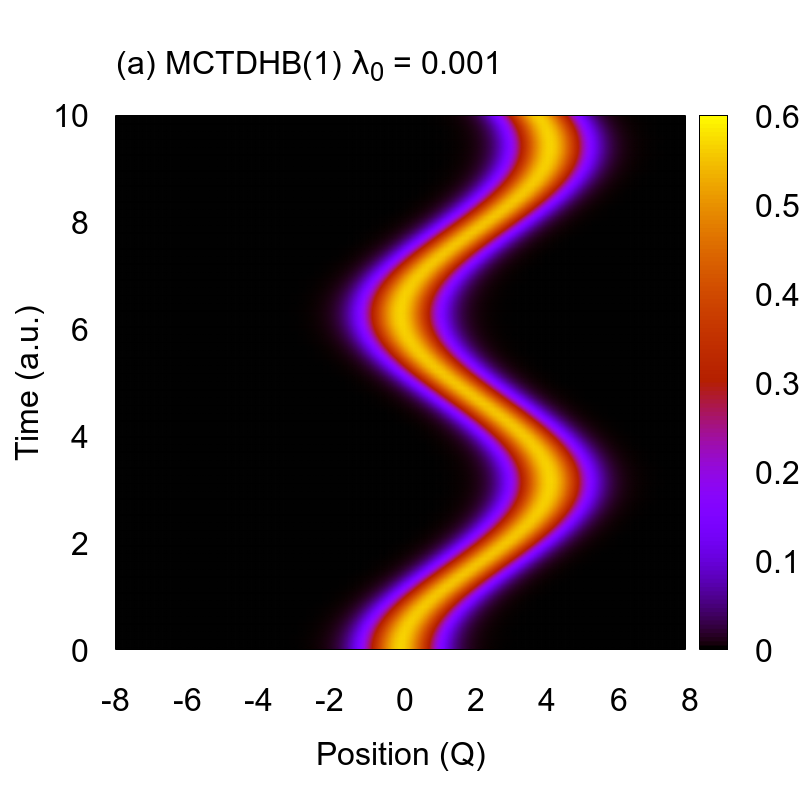}
\includegraphics[width=0.9\textwidth]{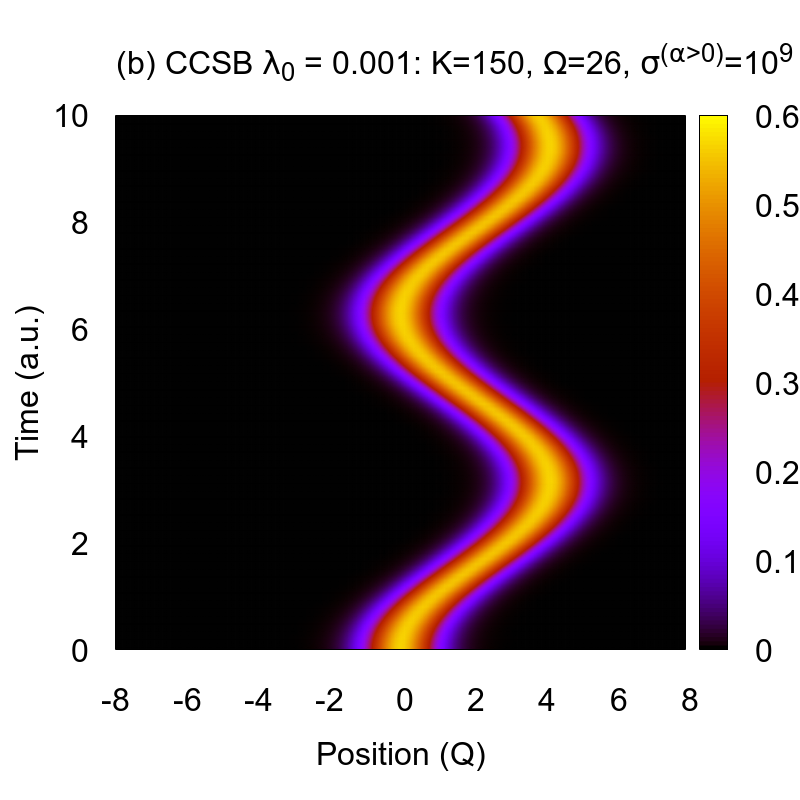}
\phantomcaption
\label{fig:density_map_lambda001}
\end{subfigure}
\begin{subfigure}{0.48\textwidth}
(ii)

\includegraphics[width=0.9\textwidth]{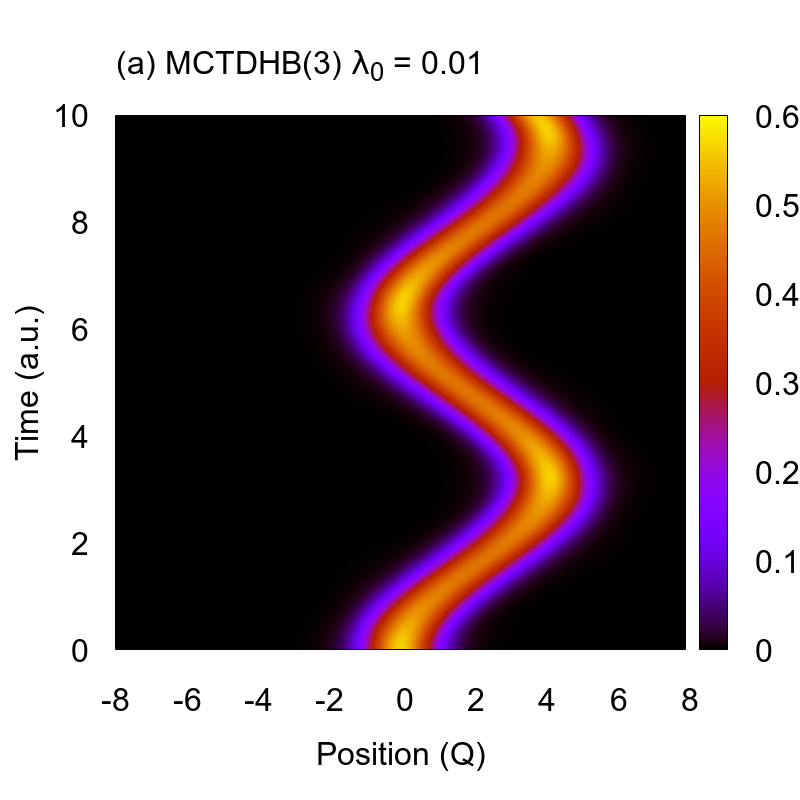}
\includegraphics[width=0.9\textwidth]{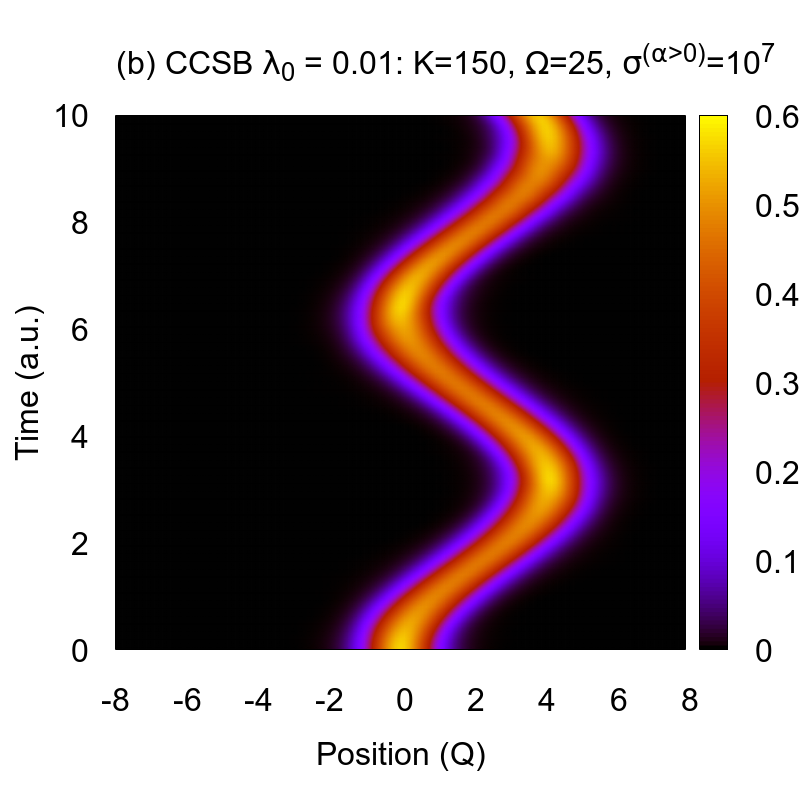}
\phantomcaption
\label{fig:density_map_lambda01}
\end{subfigure}
\caption{Space-time representation of the evolution of the 1-body density for (a) MCTDHB and (b) CCSB calculations of Application 2 with interaction strengths (i) $\lambda_0=0.001$ and (ii) $\lambda_0=0.01$.}
\label{fig:density_map}
\end{figure*}

\begin{figure*}
\centering
\begin{subfigure}{0.48\textwidth}
(i)

\includegraphics[width=0.9\textwidth]{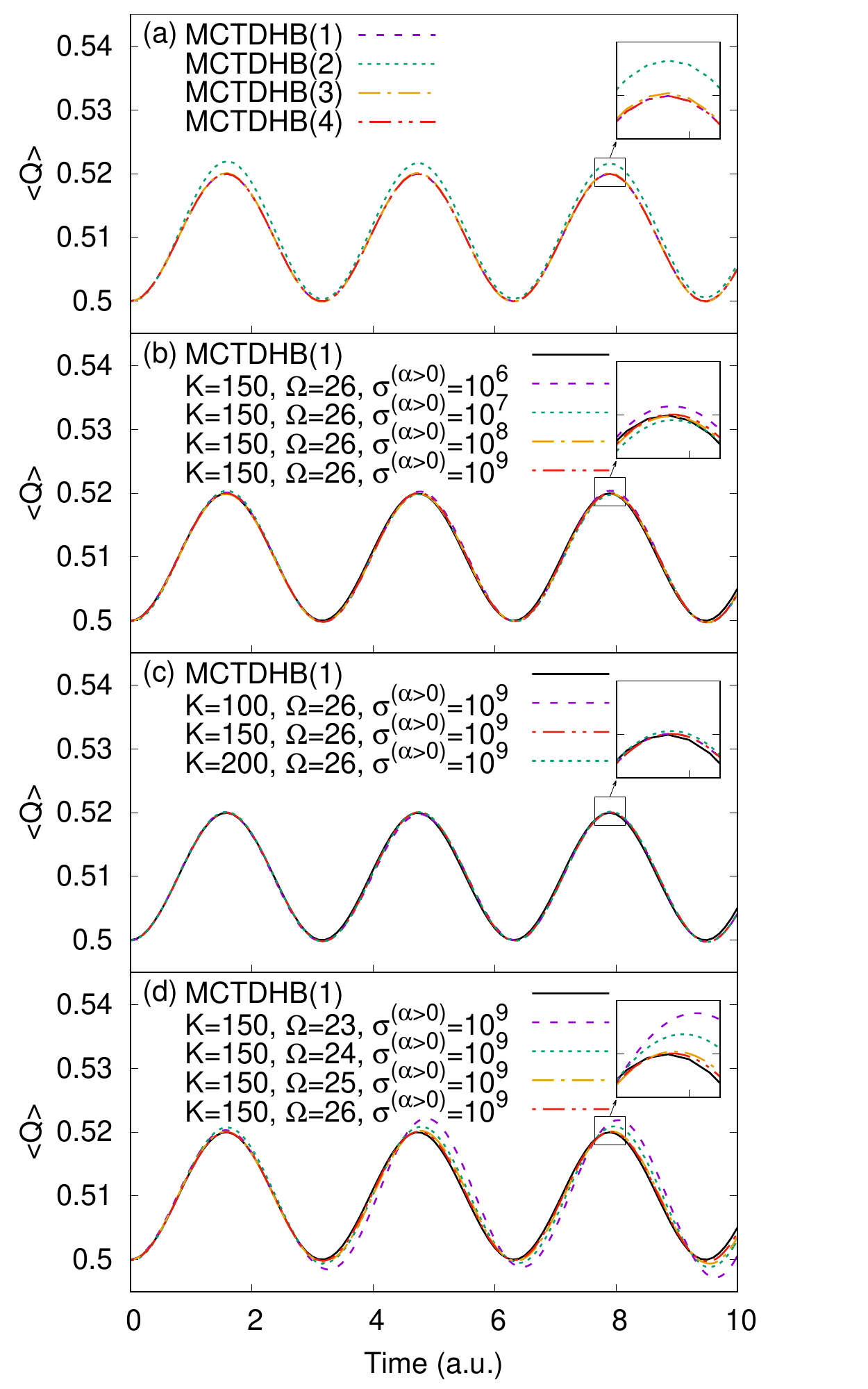}
\phantomcaption
\label{fig:density_var_lambda001}
\end{subfigure}
\begin{subfigure}{0.48\textwidth}
(ii)

\includegraphics[width=0.9\textwidth]{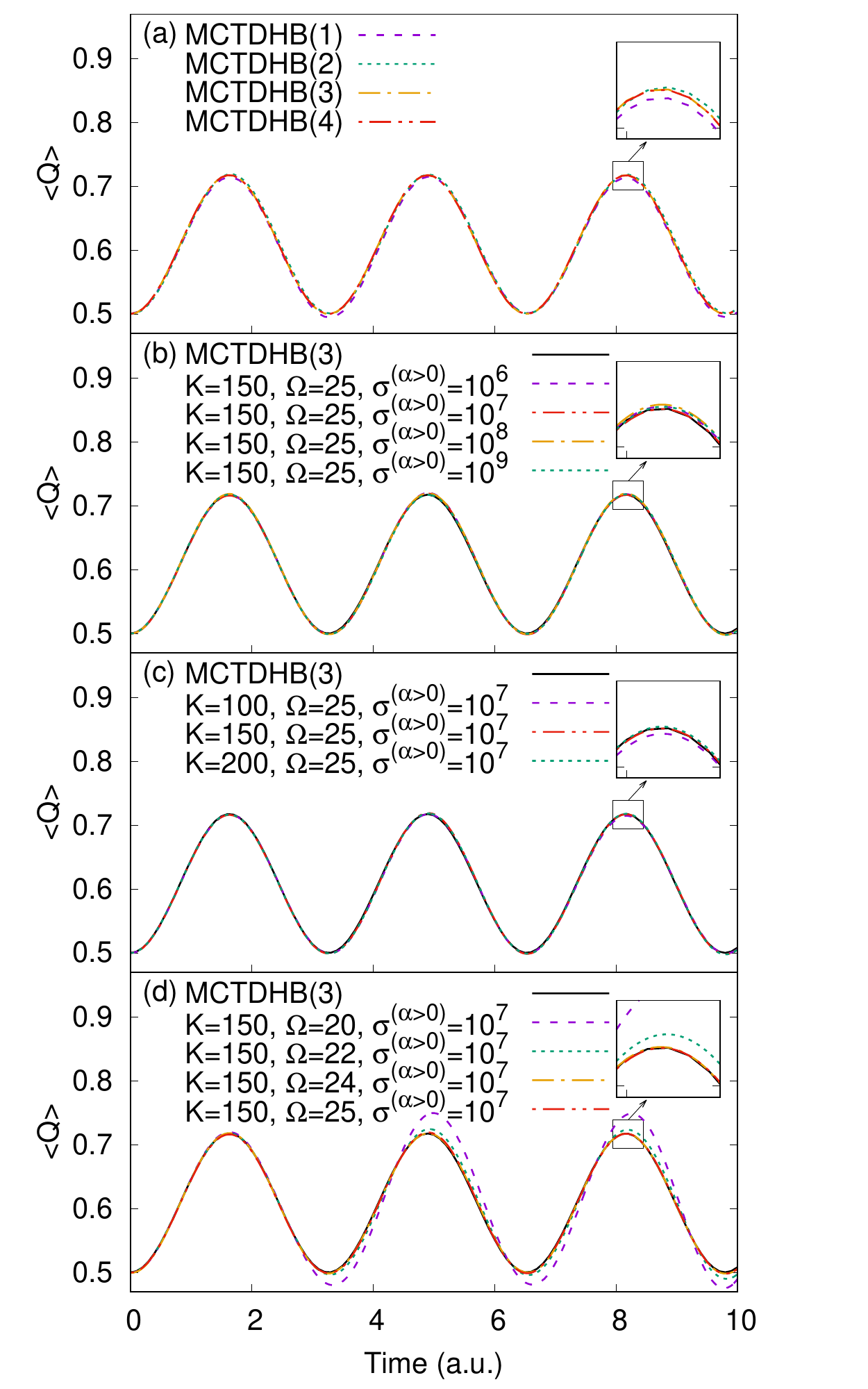}
\phantomcaption
\label{fig:density_var_lambda01}
\end{subfigure}
\caption{Variance of the 1-body density $\braket{Q}$ for Application 2 with two-body interaction strength (i) $\lambda_0=0.001$ and (ii) $\lambda_0=0.01$. In the (a) panels MCTDHB calculations with different numbers of orbitals are shown, and the converged result is used to compare to CCSB calculations with different values of (b) compression parameter for coherent state sampling of the harmonic oscillator levels with zero initial occupation $\sigma^{(\alpha>0)}$, (c) configurations $K$ and (d) harmonic oscillator levels in the basis $\Omega$.}
\label{fig:density_var}
\end{figure*}

\begin{figure*}
\centering
\begin{subfigure}{0.48\textwidth}
(i)

\includegraphics[width=0.9\textwidth]{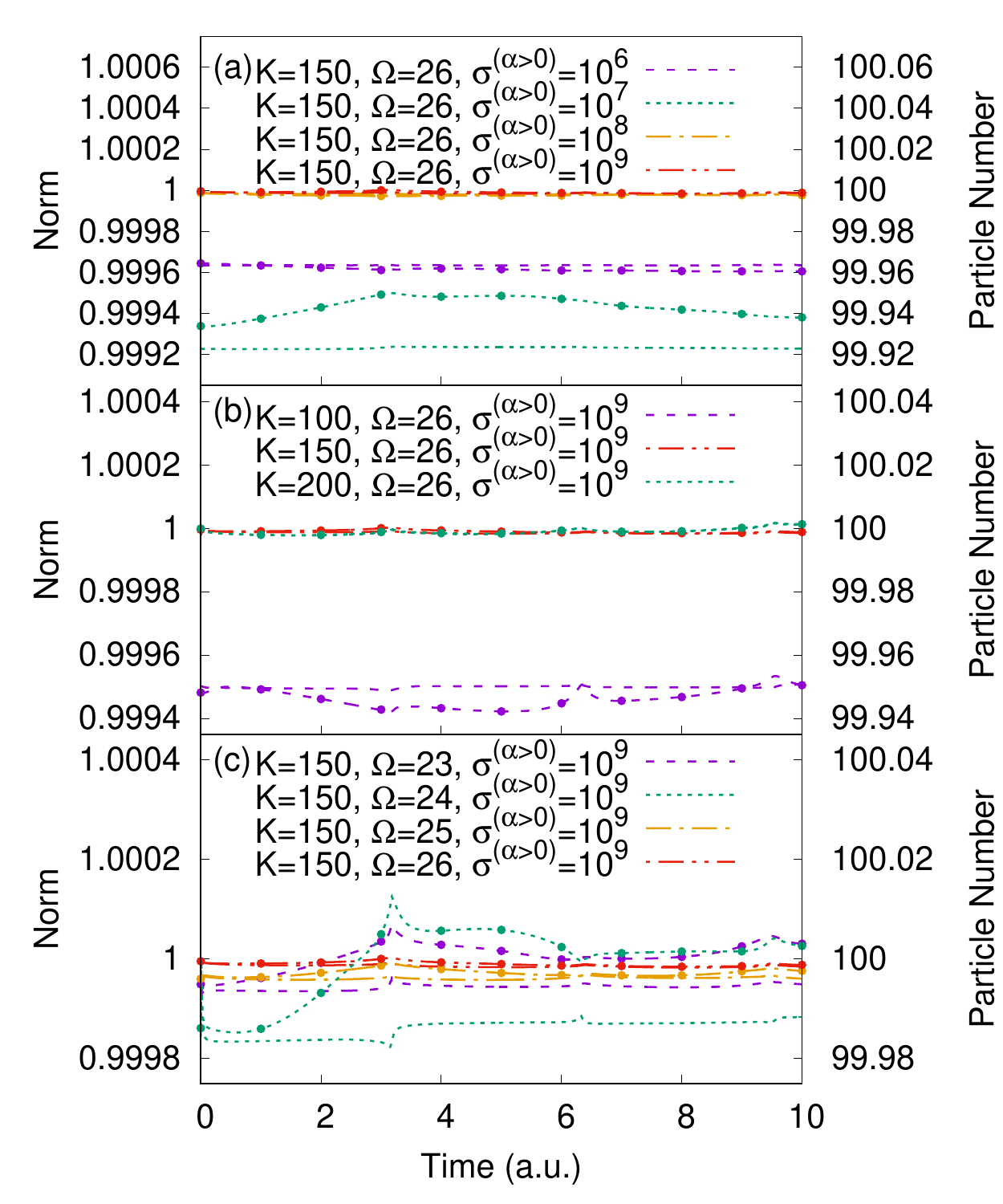}
\phantomcaption
\label{fig:norm_part_num_app2_lambda001}
\end{subfigure}
\begin{subfigure}{0.48\textwidth}
(ii)

\includegraphics[width=0.9\textwidth]{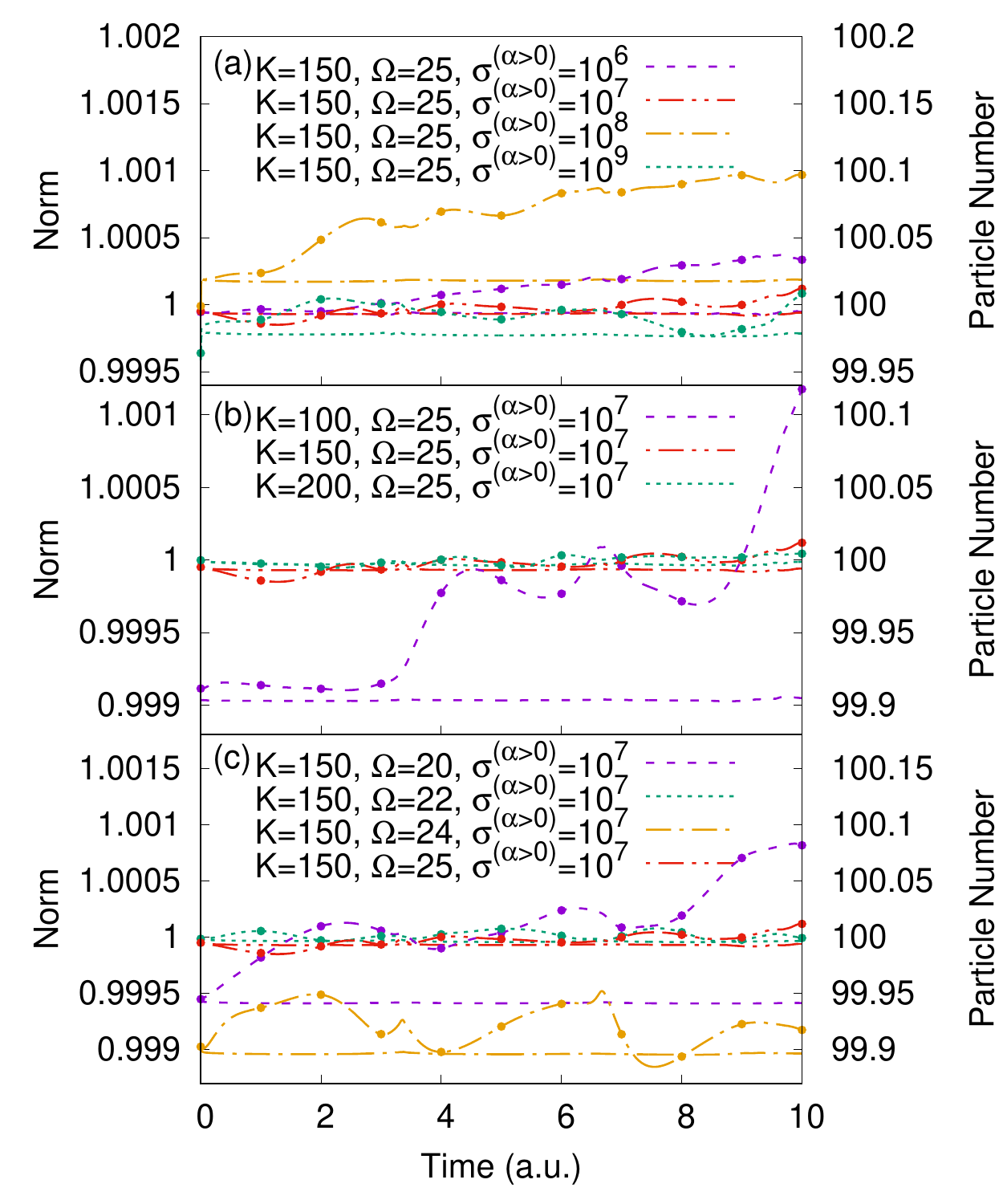}
\phantomcaption
\label{fig:norm_part_num_app2_lambda01}
\end{subfigure}
\caption{Norm (dashed/dotted lines without circles) and particle number (dashed/dotted lines with circles) for CCSB calculations of Application 2 with two-body interaction strength (i) $\lambda_0=0.001$ and (ii) $\lambda_0=0.01$ with different values of (a) compression parameter for coherent state sampling of the harmonic oscillator levels with zero initial occupation $\sigma^{(\alpha>0)}$, (b) configurations $K$ and (c) harmonic oscillator levels in the basis $\Omega$.}
\label{fig:norm_part_num_app2}
\end{figure*}

The dynamics are followed by observing the evolution of the density matrix over the course of the calculation, which in CCSB can be evaluated as
\begin{equation}\label{eq:density_Fock}
\begin{split}
\rho^{(\alpha,\beta)} &= \braket{\Psi | \hat{a}^{(\alpha)\dagger} \hat{a}^{(\beta)} | \Psi } \\&
= \sum_{k,l=1}^K D_k^* D_l \euler^{i(S_l - S_k)} \braket{\zvec_k|\zvec_l} z_k^{(\alpha)*} z_l^{(\beta)}.
\end{split}
\end{equation}
As the creation and annihilation operators have different interpretations in CCSB and MCTDHB (acting on quantum states vs orbitals), the density matrix in this form also has a different interpretation. Therefore, to compare the two methods on the same footing, the 1-body density is evaluated as a function of position, which for CCSB in this application can be calculated by the following
\begin{equation} \label{eq:density_1_body}
\begin{split}
\rho(Q) &= \braket{\alpha |\rho^{(\alpha,\beta)} |\beta} \\
& = \sum_{\alpha,\beta=0}^{\Omega} \frac{1}{\sqrt{2^\alpha \alpha!}} \left(\frac{1}{\pi}\right)^{1/4} \euler^{-Q^2 / 2} {He}_\alpha(Q) \rho^{(\alpha,\beta)}\\
& \qquad \times \frac{1}{\sqrt{2^\beta \beta!}} \left(\frac{1}{\pi}\right)^{1/4} \euler^{-Q^2 / 2} {He}_\beta(Q).
\end{split}
\end{equation}
This 1-body density is shown as a function of position and time in Fig.~\ref{fig:density_map} for interaction strengths $\lambda_0=0.001$ in (i) and $\lambda_0=0.01$ in (ii), with the MCTDHB calculations in panel (a) and CCSB calculations in panel (b). The MCTDHB calculations use 1 orbital for the $\lambda_0=0.001$ case, and 3 orbitals for the $\lambda_0=0.01$ case, labelled as MCTDHB(1) and MCTDHB(3), respectively. The CCSB calculations use $K=150$ configurations for both the $\lambda_0=0.001$ and  $\lambda_0=0.01$ cases, with $\Omega=26$ harmonic oscillator levels for the $\lambda_0=0.001$ case and  $\Omega=25$ harmonic oscillator levels for the $\lambda_0=0.01$ case, with the compression parameters  $\sigma^{(\alpha>0)}$ as mentioned in the previous section. It can be seen that the MCTDHB and CCSB calculations of the 1-body density compare well to one another, illustrating the oscillations in the bosonic cloud due to the trap displacement. 

As well as the density being subject to oscillations as a function of the trap displacement, it also exhibits breathing oscillations, which may not be immediately apparent from Fig.~\ref{fig:density_map}. To illustrate these, we plot the variance in the 1-body density as a function of time
\begin{equation}
\braket{Q} = \int Q^2 \rho(Q) \dint Q - \left(\int Q \rho(Q)  \dint Q \right)^2,
\end{equation}
which is shown in Fig.~\ref{fig:density_var}, with the interaction strengths $\lambda_0=0.001$ in (i) and $\lambda_0=0.01$ in (ii). It can immediately be seen that the stronger interaction strength produces larger breathing oscillations, which is not as easy to see in Fig.~\ref{fig:density_map}. Furthermore, the breathing oscillations  in Fig.~\ref{fig:density_var} serve to determine the convergence of the methods: MCTDHB with respect to the number of orbitals; and CCSB with respect to $\sigma^{(\alpha>0)}$, $K$, and $\Omega$. A portion of the peak of the density variance at $\sim 8$~a.u. is highlighted to clearly illustrate any discrepancies that may be difficult to distinguish. As with Application 1, the accuracy and convergence of the CCSB calculation may also be determined from the conservation of norm and particle number, which is shown in Fig.~\ref{fig:norm_part_num_app2}.

For $\lambda_0=0.001$, the MCTDHB calculations using 1, 3, and 4 orbitals demonstrate equivalent breathing oscillations in panel (a) of Fig.~\ref{fig:density_var_lambda001}, whilst there appears to be an anomalous result for MCTDHB with 2 orbitals. The authors are unsure of the reason for this, which may be best left for a future MCTDHB study, however we regard the MCTDHB(1) as being fully converged, and use this to compare to the CCSB calculations in panels (b), (c), and (d), as well as in Fig.~\ref{fig:density_map_lambda001}. An MCTDHB calculation with 1 orbital is equivalent to the GPE~\cite{sakmann2014}, so these dynamics are at the mean-field level.

In panel (b) of Fig.~\ref{fig:density_var_lambda001} we alter the value of the compression parameter $\sigma^{(\alpha>0)}$, whilst keeping the number of configurations fixed at $K=150$, and the number of harmonic oscillator levels fixed at $\Omega=26$. There appears to be little difference in the breathing oscillations for different values of $\sigma^{(\alpha>0)}$, although the highlighted peak at $\sim 8$~.a.u. demonstrates small discrepancies between the MCTDHB calculation and $\sigma^{(\alpha>0)}=10^6$ and $\sigma^{(\alpha>0)}=10^7$, whilst $\sigma^{(\alpha>0)}=10^8$ and $\sigma^{(\alpha>0)}=10^9$ superimpose on the MCTDHB result. The conservation of norm and particle number for different values of $\sigma^{(\alpha>0)}$ is shown in panel (a) of Fig.~\ref{fig:norm_part_num_app2_lambda001}, where $\sigma^{(\alpha>0)}=10^8$ and $\sigma^{(\alpha>0)}=10^9$ superimpose upon a value of the norm of 1, and particle number 100, as should be expected. We keep $\sigma^{(\alpha>0)}=10^9$ for the remaining calculations. This is much larger than the compression parameter used in Application 1, however much fewer configurations are used in this application, and the compression parameter necessary also depends upon the problem studied, and how the dynamics affect the motion of the basis. 

In panel (c) of Fig.~\ref{fig:density_var_lambda001} we alter the value of $K$ whilst keeping $\sigma^{(\alpha>0)}$ and $\Omega$ fixed. Altering the value of the compression parameter for different values of $K$ was not necessary like in Application 1, as a stable basis was able to be formed. We observe very little discrepancy between the different CCSB calculations and MCTDHB, and the norm and particle number conservation for $K=150$ and $K=200$ are very similar in panel (b) of Fig.~\ref{fig:norm_part_num_app2_lambda001}. We therefore regard $K=150$ as being fully converged.

In panel (d) of Fig.~\ref{fig:density_var_lambda001} we alter the value of $\Omega$ and keep $K$ and $\sigma^{(\alpha>0)}$ fixed. As with the above, altering the value of $\sigma^{(\alpha>0)}$ was not necessary as a stable basis was able to be formed each time. Larger discrepancies between the CCSB calculations and the MCTDHB result are seen with in this panel, indicating that the choice of $\Omega$ has the largest influence on this calculation. The results for $\Omega=25$ and $\Omega=26$ superimpose, indicating that the calculation is converged by this point. The conservation of norm and particle number for both of these calculations is also similar in panel (c) of Fig.~\ref{fig:norm_part_num_app2_lambda001}.

Turning to the larger interaction strength of  $\lambda_0=0.01$, we follow the same approach as above in determining the accuracy and convergence of the calculations. Initially, in panel (a) of Fig.~\ref{fig:density_var_lambda001} the MCTDHB calculations with 1 and 2 orbitals exhibit minor differences to those with 3 and 4 orbitals, shown in the highlighted portion of the figure, so we regard the MCTDHB(3) calculation as our fully converged reference point. At this level of interaction strength, an above mean-field description is therefore necessary. We admit that the discrepancy between these MCTDHB calculations is not very large, however a similar study in Ref.~\cite{lode2012} also illustrated minor differences in breathing dynamics for MCTDHB calculations with different numbers of orbitals.

In panel (b) of Fig.~\ref{fig:density_var_lambda01} we alter the value of the compression parameter $\sigma^{(\alpha>0)}$, whilst keeping the number of configurations fixed at $K=150$, and the number of harmonic oscillator levels fixed at $\Omega=25$. All the calculations superimpose, and there is even less discrepancy than in the $\lambda_0=0.001$ case. For the remaining calculations we choose a compression parameter of $\sigma^{(\alpha>0)}=10^7$ as this demonstrates the best norm and particle number conservation in panel (a) of  Fig.~\ref{fig:norm_part_num_app2_lambda01}.

In panel (c) of Fig.~\ref{fig:density_var_lambda01} we alter the value of $K$ whilst keeping $\sigma^{(\alpha>0)}$ and $\Omega$ fixed. There are minor differences between the $K=100$ calculation and the $K=150$ and $K=200$ calculations, the latter of which superimpose on the MCTDHB result. The norm and particle number conservation of the $K=150$ and $K=200$ calculations are very similar in panel (b) of Fig.~\ref{fig:norm_part_num_app2_lambda01}, therefore we regard the $K=150$ result as being fully converged. 

In panel (d) of Fig.~\ref{fig:density_var_lambda01} we alter the value of $\Omega$ and keep $K$ and $\sigma^{(\alpha>0)}$ fixed. As with the $\lambda_0=0.001$ calculations, this has the largest effect on the breathing oscillations, with $\Omega=20$ and $\Omega=22$ being insufficient to describe them accurately, whilst the $\Omega=24$ and $\Omega=25$ cases superimpose upon the MCTDHB result. The norm and particle number conservation of the $\Omega=24$ result, shown in panel (c) of Fig.~\ref{fig:norm_part_num_app2_lambda01} is not as good as the $\Omega=25$ result, which is why we choose the latter as our most accurate calculation.

The above demonstrates that CCSB is able to reproduce MCTDHB calculations in both the mean field and multi orbital fully quantum regimes, with similar levels of theory for the CCSB calculations in each regime. We have also shown that the method converges appropriately with respect to the $K$ and $\Omega$ parameters, this it is stable with respect to norm and particle number conservation, and that appropriate choice of the compression parameter $\sigma^{(\alpha>0)}$ for initial sampling of the coherent state basis is necessary, like in Application 1.

\section{Conclusions}

In this work the CCS method has been straightforwardly applied to investigation of indistinguishable bosons, as MCTDH and ML-MCTDH have been, and the method dubbed CCSB. Instead of the coherent state basis functions being used to represent individual particles like in the standard distinguishable representation of CCS, in CCSB they are used as a basis for number occupation of quantum states in the second quantisation Fock state formalism. 

Two example model Hamiltonians have been studied, demonstrating the accuracy of the method by comparing to fully quantum benchmarks. In the first example, CCSB was applied to the system-bath asymmetric double well tunnelling problem previously studied in Refs.~\cite{wu2004,sherratt2006,habershon2012,green2015,green2016,saller2017} in distinguishable representation. As the bath is comprised of oscillators of the same frequency, they were treated as indistinguishable and the bath portion of the Hamiltonian second quantised. The system tunnelling portion of the Hamiltonian was kept in distinguishable representation, therefore this first application was a hybrid of standard CCS and CCSB. This does not pose a problem however, as the working equations for trajectories and time-dependence of amplitudes are the same in each. This may also be thought of as a system with a bosonic bath and an impurity, opening up the possibility of the method studying multi-atomic Bose-Einstein condensates~\cite{modugno2002}, spinor Bose-Einstein condensates~\cite{kawaguchi2012}, dark-bright solitons~\cite{becker2008}, and Bose-polarons~\cite{bruderer2007}. The previously studied 20D, quadratic system-bath coupling with constant $\lambda=0.1$ case~\cite{wu2004,sherratt2006,habershon2012,green2015,green2016,saller2017} was investigated, and the second quantised bath required $\Omega=5$ harmonic oscillator levels in the basis, thus the dimensionality of the problem was reduced from 20 to 6. The CCSB calculation was in much better agreement with a benchmark result~\cite{green2015} on the system than all other methods that have studied the problem. 

In the second example, a model Hamiltonian for a system of 100 bosons in a shifted harmonic trap was studied, the 1-body density has been calculated, as well as its variance, to demonstrate the breathing oscillations of the density. Matrix elements of 2-body operators had to be calculated, as is common for interacting condensates, and these may be computed analytically by CCSB. The density oscillations were calculated at two different two-body interaction strengths and compared to MCTDHB benchmark calculations. The weaker interaction strength was able to be described by MCTDHB with 1 orbital, such that it was equivalent to the GPE mean-field theory, whilst the stronger interaction strength required MCTDHB with 3 orbitals, and was thus fully quantum, taking correlations into account and going beyond the mean-field approach. CCSB was able to reproduce both results with similar levels of theory, providing motivation for further study on more challenging Bose-Einstein condensate systems. In particular, future avenues of research for CCSB in this vein include more complicated Bose-Einstein condensate problems, such as that in Ref.~\cite{alon2008} of a condensate in a single well trap that is deformed into a double well, like that observed in experimental bosonic Josephson junctions~\cite{gati2006,gati2007}. We also wish to consider condensates in multi-well traps, such as a multi-site version of the Bose-Hubbard model studied in Ref.~\cite{ray2016}, and other interesting systems that have previously been studied by ML-MCTDHB~\cite{mistakidis2014,mistakidis2015,mistakidis2015a,mistakidis2017b,koutentakis2017,neuhaus-steinmetz2017,mistakidis2018a,plassmann2018}.

Both applications have demonstrated that the CCSB method converges with the number of configurations $K$ and number of quantum states included in the basis $\Omega$. We have also demonstrated that appropriate sampling of the initial coherent states via a compression parameter $\sigma$ is necessary to ensure a reliable and accurate calculation. Further developments of the method that we envisage in future include development of methods to generate initial conditions, as imaginary time propagation is unstable with trajectories; incorporation of SU($n$) coherent states, as demonstrated in Ref.~\citen{grigolo2016}; and the combination of the method with one to treat identical fermions~\cite{shalashilin2018} to study Bose-Fermi mixtures, as has been carried out by MCTDH~\cite{alon2007} and ML-MCTDH~\cite{cao2017} previously.

Data generated in this work that is shown in the figures may be found at \url{https://doi.org/10.5518/595}, along with the program code used to generate the data.

\section{Acknowledgements}

J.A.G. has been supported by EPSRC grant EP/N007549/1, as well as the University Research Scholarship from the University of Leeds, and funding from the School of Chemistry, University of Leeds. D.V.S acknowledges the support of the EPSRC, grant EP/P021123/1. J.A.G. would like to thank A. Streltsov for demonstration of the use of the MCTDHB program, alongside helpful discussions. J.A.G would also like to thank L. Chen for reading through the manuscript and suggesting modifications. J.A.G. and D.V.S gratefully acknowledge V. Batista, S. Habershon, and M. Saller for providing their data. This work was undertaken on ARC2 and ARC3, part of the High Performance Computing facilities at the University of Leeds, UK.

\appendix

\begin{widetext}
\section{Second Quantisation and Normal Ordering of Hamiltonian for Application 1}\label{sec:2Q_H1}

Using the definition of a second quantised Hamiltonian in Eq.~\ref{eq:2q_gen_H} in the main text, Eq.~\ref{eq:H_AS} may be written as
\begin{equation}\label{eq:2Q_H}
\begin{split}
    \hat{H}
    &= \frac{\hat{p}^{(m=1)^2}}{2} - \frac{\hat{q}^{(m=1)^2}}{2} + \frac{\hat{q}^{(m=1)^4}}{16 \eta}
    +  \left[\sum_{\alpha,\beta=0}^{\Omega} \braket{\alpha | \frac{\hat{\Pvec}^2}{2} + \frac{\hat{\Qvec}^2}{2} | \beta } \hat{a}^{(\alpha)\dagger} \hat{a}^{(\beta)}  \right] +  \frac{\lambda \hat{q}^{(m=1)}}{2} \left[ \sum_{\alpha,\beta=0}^{\Omega} \braket{\alpha | \hat{\Qvec}^2 | \beta } \hat{a}^{(\alpha)\dagger} \hat{a}^{(\beta)} \right] \\
    &= \frac{\hat{p}^{(m=1)^2}}{2} - \frac{\hat{q}^{(m=1)^2}}{2} + \frac{\hat{q}^{(m=1)^4}}{16 \eta}
    + \left[ \sum_{\alpha=0}^{\Omega}   \braket{\alpha | \frac{\hat{\Pvec}^2}{2} + \frac{\hat{\Qvec}^2}{2} | \alpha } \hat{a}^{(\alpha)\dagger} \hat{a}^{(\alpha)} \right] +  \frac{\lambda \hat{q}^{(m=1)}}{2} \left[ \sum_{\alpha,\beta=0}^{\Omega} Q^{(\alpha,\beta)^2} \hat{a}^{(\alpha)\dagger} \hat{a}^{(\beta)} \right] \\
    &= \frac{\hat{p}^{(m=1)^2}}{2} - \frac{\hat{q}^{(m=1)^2}}{2} + \frac{\hat{q}^{(m=1)^4}}{16 \eta}
    + \left[ \sum_{\alpha=0}^{\Omega} \epsilon^{(\alpha)} \hat{a}^{(\alpha)\dagger} \hat{a}^{(\alpha)}   \right] +  \frac{\lambda \hat{q}^{(m=1)}}{2} \left[ \sum_{\alpha,\beta=0}^{\Omega} Q^{(\alpha,\beta)^2}  \hat{a}^{(\alpha)\dagger} \hat{a}^{(\beta)} \right].
\end{split}
\end{equation}
The quantum states $\ket{\alpha}$ and $\ket{\beta}$ are those of the harmonic oscillator with $\alpha$ and $\beta$ numbers of quanta, and the equality on the second line for $\braket{\alpha | \frac{\hat{\Pvec}^2}{2} + \frac{\hat{\Qvec}^2}{2} | \beta }$ follows because this is non-zero with eigenvalue $\epsilon^{(\alpha)}$ only when $\alpha=\beta$. The sums are from the ground level $\alpha=0$, to some upper level $\Omega$. In principle, one should choose $\Omega = \infty$ for a complete description of the bath, however in practice additional oscillator levels may simply be added on until a converged result is achieved. The position and momentum operators of the tunnelling mode have explicitly been labelled with $(m=1)$ to distinguish them from the $\alpha$ labelling scheme of the second quantised bath modes.

The matrix $Q^{(\alpha,\beta)^2}$ is evaluated as
\begin{equation}\label{eq:Q2_mat_el}
\braket{\alpha| \hat{\Qvec}^2 | \beta} = \begin{dcases}
     \frac{1}{2} \sqrt{(\alpha+2)(\alpha+1)} & \text{if } \alpha=\beta-2 \\
     \frac{1}{2} \sqrt{\alpha(\alpha-1)} & \text{if } \alpha=\beta+2 \\
     \epsilon^{(\alpha)} & \text{if } \alpha = \beta \\
     0 & \text{otherwise.}
    \end{dcases}
\end{equation}
As this matrix is non-zero only for quanta $\alpha=\beta$ and $\alpha=\beta \pm 2$, and we may say that all bath modes are initially in the ground harmonic oscillator level ($\alpha=0$) as they are at the origin in distinguishable representation (previously assumed in the benchmark calculation~\cite{green2015}), only harmonic oscillator levels with even numbers of quanta will be included and the bottom line of Eq.~\ref{eq:2Q_H} is written as
\begin{equation}\label{eq:2Q_H2}
\hat{H} =  \frac{\hat{p}^{(m=1)^2}}{2} - \frac{\hat{q}^{(m=1)^2}}{2} + \frac{\hat{q}^{(m=1)^4}}{16 \eta}
    + \left[ \sum_{\alpha=0}^{\Omega} \epsilon^{(2\alpha)} \hat{a}^{(2\alpha)\dagger} \hat{a}^{(2\alpha)}   \right] +  \frac{\lambda \hat{q}^{(m=1)}}{2} \left[ \sum_{\alpha,\beta=0}^{\Omega} Q^{(2\alpha,2\beta)^2}  \hat{a}^{(2\alpha)\dagger} \hat{a}^{(2\beta)} \right].
\end{equation}
The relationship between the creation and annihilation operators and $\hat{q}$ and $\hat{p}$ given in Eq.~\ref{eq:creat_ann} may then be used in Eq.~\ref{eq:2Q_H2} alongside the relationships in Eqs.~\ref{eq:creat_ann_eig} and~\ref{eq:H_CS} to give Eq.~\ref{eq:Hord_1}.

\section{Second Quantisation of Hamiltonian for Application 2}\label{sec:2Q_H2}

Using the definition of a second quantised Hamiltonian in Eq.~\ref{eq:2q_gen_H} in the main text, Eq.~\ref{eq:H_BC} may be written as
\begin{equation}
\label{eq:H_BC_2Q}
\begin{split}
\hat{H} =& \sum_{\alpha,\beta=0}^{\Omega} \braket{\alpha | \frac{\hat{\Pvec}^2}{2} + \frac{\hat{\Qvec}^2}{2} | \beta} \hat{a}^{(\alpha)\dagger} \hat{a}^{(\beta)}  - \sum_{\alpha,\beta=0}^{\Omega} \braket{\alpha | \xi \hat{\Qvec} | \beta} \hat{a}^{(\alpha)\dagger} \hat{a}^{(\beta)}  + \sum_{\alpha,\beta=0}^{\Omega} \braket{\alpha| \frac{\xi^2}{2} |\beta} \hat{a}^{(\alpha)\dagger} \hat{a}^{(\beta)} \\
& + \frac{1}{2} \sum_{\alpha,\beta,\gamma,\zeta=0}^{\Omega} \braket{\alpha,\beta | \lambda_0 \delta(\Qvec-\Qvec') | \gamma, \zeta} \hat{a}^{(\alpha)\dagger} \hat{a}^{(\beta)\dagger} \hat{a}^{(\zeta)} \hat{a}^{(\gamma)}  \\
=& \sum_{\alpha=0}^{\Omega} \braket{\alpha | \frac{\hat{\Pvec}^2}{2} + \frac{\hat{\Qvec}^2}{2} | \alpha} \hat{a}^{(\alpha)\dagger} \hat{a}^{(\alpha)}  - \sum_{\alpha,\beta=0}^{\Omega} \braket{\alpha | \xi \hat{\Qvec} | \beta} \hat{a}^{(\alpha)\dagger} \hat{a}^{(\beta)} + \sum_{\alpha=0}^{\Omega} \braket{\alpha| \frac{\xi^2}{2} |\alpha} \hat{a}^{(\alpha)\dagger} \hat{a}^{(\alpha)} \\
& + \frac{1}{2} \sum_{\alpha,\beta,\gamma,\zeta=0}^{\Omega} \braket{\alpha,\beta | \lambda_0 \delta(\Qvec-\Qvec') | \gamma, \zeta} \hat{a}^{(\alpha)\dagger} \hat{a}^{(\beta)\dagger} \hat{a}^{(\zeta)} \hat{a}^{(\gamma)} \\
=& \sum_{\alpha=0}^{\Omega} \epsilon^{(\alpha)} \hat{a}^{(\alpha)\dagger} \hat{a}^{(\alpha)} - \sum_{\alpha,\beta=0}^{\Omega}  \xi Q^{(\alpha,\beta)} \hat{a}^{(\alpha)\dagger} \hat{a}^{(\beta)} + \sum_{\alpha=0}^{\Omega} \frac{\xi^2}{2} \hat{a}^{(\alpha)\dagger} \hat{a}^{(\alpha)} + \frac{1}{2} \sum_{\alpha,\beta,\gamma,\zeta=0}^{\Omega} \lambda_0 \delta^{(\alpha,\beta,\gamma,\zeta)} \hat{a}^{(\alpha)\dagger} \hat{a}^{(\beta)\dagger} \hat{a}^{(\zeta)} \hat{a}^{(\gamma)}.
\end{split}
\end{equation}
The relationships in Eqs.~\ref{eq:creat_ann_eig} and~\ref{eq:H_CS} may then be used with Eq.~\ref{eq:H_BC_2Q} to give Eq.~\ref{eq:Hord_2}. In Eq.~\ref{eq:H_BC_2Q}, $\epsilon^{(\alpha)}$ is the eigenvalue of the harmonic oscillator for state $\ket{\alpha}$, and $Q^{(\alpha,\beta)}$ is a matrix given by
\begin{equation}
Q^{(\alpha,\beta)} = \braket{\alpha | \hat{\Qvec} | \beta} = 
\begin{cases}
 \sqrt{\frac{\alpha}{2}} & \alpha = \beta+1 \\
 \sqrt{\frac{\beta}{2}} & \beta = \alpha+1 \\
 0                  & \text{otherwise.}
\end{cases}.
\end{equation}
Evaluation of the $\delta^{(\alpha,\beta,\gamma,\zeta)}$ matrix is slightly more involved, as it is required to solve the integral
\begin{equation}
\begin{split}
\delta^{(\alpha,\beta,\gamma,\zeta)} =& \braket{\alpha , \beta | \delta(\Qvec-\Qvec') | \gamma , \zeta} \\
=& \int_{-\infty}^{+\infty} \int_{-\infty}^{+\infty} \frac{1}{\sqrt{2^\alpha \alpha!}} \left(\frac{1}{\pi}\right)^{1/4} \euler^{-\Qvec^2/2} {He}^{(\alpha)}(\Qvec) \frac{1}{\sqrt{2^\beta \beta!}} \left(\frac{1}{\pi}\right)^{1/4} \euler^{-\Qvec'^2/2} {He}^{(\beta)}(\Qvec') \\
& \times \delta(\Qvec-\Qvec') \frac{1}{\sqrt{2^\gamma \gamma!}} \left(\frac{1}{\pi}\right)^{1/4} \euler^{-\Qvec^2/2} {He}^{(\gamma)}(\Qvec) \frac{1}{\sqrt{2^\zeta \zeta!}} \left(\frac{1}{\pi}\right)^{1/4} \euler^{-\Qvec'^2/2} {He}^{(\zeta)}(\Qvec') \dint \Qvec \dint \Qvec'
\end{split}
\end{equation}
where ${He}^{(\alpha)}(\Qvec)$ is a Hermite polynomial of order $\alpha$. However, an analytic solution is possible, and the above may be simplified using the relationship
\begin{equation}
\int_{-\infty}^{+\infty} f(x') \delta(x-x') \dint x' = f(x),
\end{equation}
and like terms collated to obtain
\begin{equation}
\delta^{(\alpha,\beta,\gamma,\zeta)} = \frac{1}{\pi \sqrt{2^{(\alpha+\beta+\gamma+\zeta)} \alpha! \beta! \gamma! \zeta!}} \int_{-\infty}^{+\infty}  \euler^{-2\Qvec^2} {He}^{(\alpha)}(\Qvec) {He}^{(\beta)}(\Qvec) {He}^{(\gamma)}(\Qvec) {He}^{(\zeta)}(\Qvec) \dint \Qvec.
\end{equation}
This will only be non-zero if the integrand is an even function, so the product of Hermite polynomials can only have even powers of $\Qvec$
\begin{equation}\label{eq:delta_int1}
\delta^{(\alpha,\beta,\gamma,\zeta)} = \frac{1}{\pi \sqrt{2^{\alpha+\beta+\gamma+\zeta} \alpha! \beta! \gamma! \zeta!}} \int_{-\infty}^{+\infty} \euler^{-2\Qvec^2} \sum_{\tau=0}^{(\alpha+\beta+\gamma+\zeta)/2} c_{2\tau} \Qvec^{2\tau} \dint \Qvec
\end{equation}
where $c_{2\tau}$ is a constant obtained from the product of Hermite polynomial coefficients. Using the following identity
\begin{equation}
\int_{-\infty}^{+\infty} x^{2n} \euler^{-\frac{1}{2} a x^2} \dint x = \sqrt{\frac{2\pi}{a}} \frac{1}{a^n} (2n-1)!! \quad \text{for}\, n>0
\end{equation}
combined with a Gaussian integral for $\tau=0$, Eq.~\ref{eq:delta_int1} can be evaluated as
\begin{equation}\label{eq:delta_mat_el}
\delta^{(\alpha,\beta,\gamma,\zeta)} = \frac{1}{\pi \sqrt{2^{\alpha+\beta+\gamma+\zeta} \alpha! \beta! \gamma! \zeta!}} \left[ \sqrt{\frac{\pi}{2}}c_0 + \sum_{\tau=1}^{(\alpha+\beta+\gamma+\zeta)/2} c_{2\tau} \sqrt{\frac{\pi}{2}} \frac{1}{4^\tau} (2\tau-1)!! \right] .
\end{equation}

We note that an alternative method of calculating the $\delta^{(\alpha,\beta,\gamma,\zeta)}$ matrix elements exists using Gauss-Hermite quadrature~\cite{edwards1996}, however our approach was sufficiently efficient.

\end{widetext}

%\section{References}

\bibliographystyle{apsrev4-1}

%\bibliographystyle{bibstyle}
%\bibliography{ref_list}

\begin{thebibliography}{87}%
\makeatletter
\providecommand \@ifxundefined [1]{%
 \@ifx{#1\undefined}
}%
\providecommand \@ifnum [1]{%
 \ifnum #1\expandafter \@firstoftwo
 \else \expandafter \@secondoftwo
 \fi
}%
\providecommand \@ifx [1]{%
 \ifx #1\expandafter \@firstoftwo
 \else \expandafter \@secondoftwo
 \fi
}%
\providecommand \natexlab [1]{#1}%
\providecommand \enquote  [1]{``#1''}%
\providecommand \bibnamefont  [1]{#1}%
\providecommand \bibfnamefont [1]{#1}%
\providecommand \citenamefont [1]{#1}%
\providecommand \href@noop [0]{\@secondoftwo}%
\providecommand \href [0]{\begingroup \@sanitize@url \@href}%
\providecommand \@href[1]{\@@startlink{#1}\@@href}%
\providecommand \@@href[1]{\endgroup#1\@@endlink}%
\providecommand \@sanitize@url [0]{\catcode `\\12\catcode `\$12\catcode
  `\&12\catcode `\#12\catcode `\^12\catcode `\_12\catcode `\%12\relax}%
\providecommand \@@startlink[1]{}%
\providecommand \@@endlink[0]{}%
\providecommand \url  [0]{\begingroup\@sanitize@url \@url }%
\providecommand \@url [1]{\endgroup\@href {#1}{\urlprefix }}%
\providecommand \urlprefix  [0]{URL }%
\providecommand \Eprint [0]{\href }%
\providecommand \doibase [0]{http://dx.doi.org/}%
\providecommand \selectlanguage [0]{\@gobble}%
\providecommand \bibinfo  [0]{\@secondoftwo}%
\providecommand \bibfield  [0]{\@secondoftwo}%
\providecommand \translation [1]{[#1]}%
\providecommand \BibitemOpen [0]{}%
\providecommand \bibitemStop [0]{}%
\providecommand \bibitemNoStop [0]{.\EOS\space}%
\providecommand \EOS [0]{\spacefactor3000\relax}%
\providecommand \BibitemShut  [1]{\csname bibitem#1\endcsname}%
\let\auto@bib@innerbib\@empty
%</preamble>
\bibitem [{\citenamefont {Anderson}\ \emph {et~al.}(1995)\citenamefont
  {Anderson}, \citenamefont {Ensher}, \citenamefont {Matthews}, \citenamefont
  {Wieman},\ and\ \citenamefont {Cornell}}]{anderson1995}%
  \BibitemOpen
  \bibfield  {author} {\bibinfo {author} {\bibfnamefont {M.~H.}\ \bibnamefont
  {Anderson}}, \bibinfo {author} {\bibfnamefont {J.~R.}\ \bibnamefont
  {Ensher}}, \bibinfo {author} {\bibfnamefont {M.~R.}\ \bibnamefont
  {Matthews}}, \bibinfo {author} {\bibfnamefont {C.~E.}\ \bibnamefont
  {Wieman}}, \ and\ \bibinfo {author} {\bibfnamefont {E.~A.}\ \bibnamefont
  {Cornell}},\ }\href {\doibase 10.1126/science.269.5221.198} {\bibfield
  {journal} {\bibinfo  {journal} {Science}\ }\textbf {\bibinfo {volume}
  {269}},\ \bibinfo {pages} {198} (\bibinfo {year} {1995})}\BibitemShut
  {NoStop}%
\bibitem [{\citenamefont {Bradley}\ \emph {et~al.}(1995)\citenamefont
  {Bradley}, \citenamefont {Sackett}, \citenamefont {Tollett},\ and\
  \citenamefont {Hulet}}]{bradley1995}%
  \BibitemOpen
  \bibfield  {author} {\bibinfo {author} {\bibfnamefont {C.~C.}\ \bibnamefont
  {Bradley}}, \bibinfo {author} {\bibfnamefont {C.~A.}\ \bibnamefont
  {Sackett}}, \bibinfo {author} {\bibfnamefont {J.~J.}\ \bibnamefont
  {Tollett}}, \ and\ \bibinfo {author} {\bibfnamefont {R.~G.}\ \bibnamefont
  {Hulet}},\ }\href {\doibase 10.1103/PhysRevLett.75.1687} {\bibfield
  {journal} {\bibinfo  {journal} {Physical Review Letters}\ }\textbf {\bibinfo
  {volume} {75}},\ \bibinfo {pages} {1687} (\bibinfo {year}
  {1995})}\BibitemShut {NoStop}%
\bibitem [{\citenamefont {Davis}\ \emph {et~al.}(1995)\citenamefont {Davis},
  \citenamefont {Mewes}, \citenamefont {Andrews}, \citenamefont {van Druten},
  \citenamefont {Durfee}, \citenamefont {Kurn},\ and\ \citenamefont
  {Ketterle}}]{davis1995}%
  \BibitemOpen
  \bibfield  {author} {\bibinfo {author} {\bibfnamefont {K.~B.}\ \bibnamefont
  {Davis}}, \bibinfo {author} {\bibfnamefont {M.~O.}\ \bibnamefont {Mewes}},
  \bibinfo {author} {\bibfnamefont {M.~R.}\ \bibnamefont {Andrews}}, \bibinfo
  {author} {\bibfnamefont {N.~J.}\ \bibnamefont {van Druten}}, \bibinfo
  {author} {\bibfnamefont {D.~S.}\ \bibnamefont {Durfee}}, \bibinfo {author}
  {\bibfnamefont {D.~M.}\ \bibnamefont {Kurn}}, \ and\ \bibinfo {author}
  {\bibfnamefont {W.}~\bibnamefont {Ketterle}},\ }\href {\doibase
  10.1103/PhysRevLett.75.3969} {\bibfield  {journal} {\bibinfo  {journal}
  {Physical Review Letters}\ }\textbf {\bibinfo {volume} {75}},\ \bibinfo
  {pages} {3969} (\bibinfo {year} {1995})}\BibitemShut {NoStop}%
\bibitem [{\citenamefont {Orzel}\ \emph {et~al.}(2001)\citenamefont {Orzel},
  \citenamefont {Tuchman}, \citenamefont {Fenselau}, \citenamefont {Yasuda},\
  and\ \citenamefont {Kasevich}}]{orzel2001}%
  \BibitemOpen
  \bibfield  {author} {\bibinfo {author} {\bibfnamefont {C.}~\bibnamefont
  {Orzel}}, \bibinfo {author} {\bibfnamefont {A.~K.}\ \bibnamefont {Tuchman}},
  \bibinfo {author} {\bibfnamefont {M.~L.}\ \bibnamefont {Fenselau}}, \bibinfo
  {author} {\bibfnamefont {M.}~\bibnamefont {Yasuda}}, \ and\ \bibinfo {author}
  {\bibfnamefont {M.~A.}\ \bibnamefont {Kasevich}},\ }\href {\doibase
  10.1126/science.1058149} {\bibfield  {journal} {\bibinfo  {journal}
  {Science}\ }\textbf {\bibinfo {volume} {291}},\ \bibinfo {pages} {2386}
  (\bibinfo {year} {2001})}\BibitemShut {NoStop}%
\bibitem [{\citenamefont {Albiez}\ \emph {et~al.}(2005)\citenamefont {Albiez},
  \citenamefont {Gati}, \citenamefont {F\"olling}, \citenamefont {Hunsmann},
  \citenamefont {Cristiani},\ and\ \citenamefont {Oberthaler}}]{albiez2005}%
  \BibitemOpen
  \bibfield  {author} {\bibinfo {author} {\bibfnamefont {M.}~\bibnamefont
  {Albiez}}, \bibinfo {author} {\bibfnamefont {R.}~\bibnamefont {Gati}},
  \bibinfo {author} {\bibfnamefont {J.}~\bibnamefont {F\"olling}}, \bibinfo
  {author} {\bibfnamefont {S.}~\bibnamefont {Hunsmann}}, \bibinfo {author}
  {\bibfnamefont {M.}~\bibnamefont {Cristiani}}, \ and\ \bibinfo {author}
  {\bibfnamefont {M.~K.}\ \bibnamefont {Oberthaler}},\ }\href {\doibase
  10.1103/PhysRevLett.95.010402} {\bibfield  {journal} {\bibinfo  {journal}
  {Physical Review Letters}\ }\textbf {\bibinfo {volume} {95}},\ \bibinfo {eid}
  {010402} (\bibinfo {year} {2005})}\BibitemShut {NoStop}%
\bibitem [{\citenamefont {Levy}\ \emph {et~al.}(2007)\citenamefont {Levy},
  \citenamefont {Lahoud}, \citenamefont {Shomroni},\ and\ \citenamefont
  {Steinhauer}}]{levy2007}%
  \BibitemOpen
  \bibfield  {author} {\bibinfo {author} {\bibfnamefont {S.}~\bibnamefont
  {Levy}}, \bibinfo {author} {\bibfnamefont {E.}~\bibnamefont {Lahoud}},
  \bibinfo {author} {\bibfnamefont {I.}~\bibnamefont {Shomroni}}, \ and\
  \bibinfo {author} {\bibfnamefont {J.}~\bibnamefont {Steinhauer}},\ }\href
  {\doibase https://doi.org/10.1038/nature06186} {\bibfield  {journal}
  {\bibinfo  {journal} {Nature}\ }\textbf {\bibinfo {volume} {449}},\ \bibinfo
  {pages} {579} (\bibinfo {year} {2007})}\BibitemShut {NoStop}%
\bibitem [{\citenamefont {Matthews}\ \emph {et~al.}(1999)\citenamefont
  {Matthews}, \citenamefont {Anderson}, \citenamefont {Haljan}, \citenamefont
  {Hall}, \citenamefont {Wieman},\ and\ \citenamefont
  {Cornell}}]{matthews1999}%
  \BibitemOpen
  \bibfield  {author} {\bibinfo {author} {\bibfnamefont {M.~R.}\ \bibnamefont
  {Matthews}}, \bibinfo {author} {\bibfnamefont {B.~P.}\ \bibnamefont
  {Anderson}}, \bibinfo {author} {\bibfnamefont {P.~C.}\ \bibnamefont
  {Haljan}}, \bibinfo {author} {\bibfnamefont {D.~S.}\ \bibnamefont {Hall}},
  \bibinfo {author} {\bibfnamefont {C.~E.}\ \bibnamefont {Wieman}}, \ and\
  \bibinfo {author} {\bibfnamefont {E.~A.}\ \bibnamefont {Cornell}},\ }\href
  {\doibase 10.1103/PhysRevLett.83.2498} {\bibfield  {journal} {\bibinfo
  {journal} {Physical Review Letters}\ }\textbf {\bibinfo {volume} {83}},\
  \bibinfo {pages} {2498} (\bibinfo {year} {1999})}\BibitemShut {NoStop}%
\bibitem [{\citenamefont {Madison}\ \emph {et~al.}(2000)\citenamefont
  {Madison}, \citenamefont {Chevy}, \citenamefont {Wohlleben},\ and\
  \citenamefont {Dalibard}}]{madison2000}%
  \BibitemOpen
  \bibfield  {author} {\bibinfo {author} {\bibfnamefont {K.~W.}\ \bibnamefont
  {Madison}}, \bibinfo {author} {\bibfnamefont {F.}~\bibnamefont {Chevy}},
  \bibinfo {author} {\bibfnamefont {W.}~\bibnamefont {Wohlleben}}, \ and\
  \bibinfo {author} {\bibfnamefont {J.}~\bibnamefont {Dalibard}},\ }\href
  {\doibase 10.1103/PhysRevLett.84.806} {\bibfield  {journal} {\bibinfo
  {journal} {Physical Review Letters}\ }\textbf {\bibinfo {volume} {84}},\
  \bibinfo {pages} {806} (\bibinfo {year} {2000})}\BibitemShut {NoStop}%
\bibitem [{\citenamefont {Burger}\ \emph {et~al.}(1999)\citenamefont {Burger},
  \citenamefont {Bongs}, \citenamefont {Dettmer}, \citenamefont {Ertmer},
  \citenamefont {Sengstock}, \citenamefont {Sanpera}, \citenamefont
  {Shlyapnikov},\ and\ \citenamefont {Lewenstein}}]{burger1999}%
  \BibitemOpen
  \bibfield  {author} {\bibinfo {author} {\bibfnamefont {S.}~\bibnamefont
  {Burger}}, \bibinfo {author} {\bibfnamefont {K.}~\bibnamefont {Bongs}},
  \bibinfo {author} {\bibfnamefont {S.}~\bibnamefont {Dettmer}}, \bibinfo
  {author} {\bibfnamefont {W.}~\bibnamefont {Ertmer}}, \bibinfo {author}
  {\bibfnamefont {K.}~\bibnamefont {Sengstock}}, \bibinfo {author}
  {\bibfnamefont {A.}~\bibnamefont {Sanpera}}, \bibinfo {author} {\bibfnamefont
  {G.~V.}\ \bibnamefont {Shlyapnikov}}, \ and\ \bibinfo {author} {\bibfnamefont
  {M.}~\bibnamefont {Lewenstein}},\ }\href {\doibase
  10.1103/PhysRevLett.83.5198} {\bibfield  {journal} {\bibinfo  {journal}
  {Physical Review Letters}\ }\textbf {\bibinfo {volume} {83}},\ \bibinfo
  {pages} {5198} (\bibinfo {year} {1999})}\BibitemShut {NoStop}%
\bibitem [{\citenamefont {Denschlag}\ \emph {et~al.}(2000)\citenamefont
  {Denschlag}, \citenamefont {Simsarian}, \citenamefont {Feder}, \citenamefont
  {Clark}, \citenamefont {Collins}, \citenamefont {Cubizolles}, \citenamefont
  {Deng}, \citenamefont {Hagley}, \citenamefont {Helmerson}, \citenamefont
  {Reinhardt}, \citenamefont {Rolston}, \citenamefont {Schneider},\ and\
  \citenamefont {Phillips}}]{denschlag2000}%
  \BibitemOpen
  \bibfield  {author} {\bibinfo {author} {\bibfnamefont {J.}~\bibnamefont
  {Denschlag}}, \bibinfo {author} {\bibfnamefont {J.~E.}\ \bibnamefont
  {Simsarian}}, \bibinfo {author} {\bibfnamefont {D.~L.}\ \bibnamefont
  {Feder}}, \bibinfo {author} {\bibfnamefont {C.~W.}\ \bibnamefont {Clark}},
  \bibinfo {author} {\bibfnamefont {L.~A.}\ \bibnamefont {Collins}}, \bibinfo
  {author} {\bibfnamefont {J.}~\bibnamefont {Cubizolles}}, \bibinfo {author}
  {\bibfnamefont {L.}~\bibnamefont {Deng}}, \bibinfo {author} {\bibfnamefont
  {E.~W.}\ \bibnamefont {Hagley}}, \bibinfo {author} {\bibfnamefont
  {K.}~\bibnamefont {Helmerson}}, \bibinfo {author} {\bibfnamefont {W.~P.}\
  \bibnamefont {Reinhardt}}, \bibinfo {author} {\bibfnamefont {S.~L.}\
  \bibnamefont {Rolston}}, \bibinfo {author} {\bibfnamefont {B.~I.}\
  \bibnamefont {Schneider}}, \ and\ \bibinfo {author} {\bibfnamefont {W.~D.}\
  \bibnamefont {Phillips}},\ }\href {\doibase 10.1126/science.287.5450.97}
  {\bibfield  {journal} {\bibinfo  {journal} {Science}\ }\textbf {\bibinfo
  {volume} {287}},\ \bibinfo {pages} {97} (\bibinfo {year} {2000})}\BibitemShut
  {NoStop}%
\bibitem [{\citenamefont {Gross}(1961)}]{gross1961}%
  \BibitemOpen
  \bibfield  {author} {\bibinfo {author} {\bibfnamefont {E.~P.}\ \bibnamefont
  {Gross}},\ }\href {\doibase 10.1007/BF02731494} {\bibfield  {journal}
  {\bibinfo  {journal} {Nuovo Cimento}\ }\textbf {\bibinfo {volume} {20}},\
  \bibinfo {pages} {454} (\bibinfo {year} {1961})}\BibitemShut {NoStop}%
\bibitem [{\citenamefont {Pitaevskii}(1961)}]{pitaevskii1961}%
  \BibitemOpen
  \bibfield  {author} {\bibinfo {author} {\bibfnamefont {L.~P.}\ \bibnamefont
  {Pitaevskii}},\ }\href@noop {} {\bibfield  {journal} {\bibinfo  {journal}
  {Soviet Physics JETP-USSR}\ }\textbf {\bibinfo {volume} {13}},\ \bibinfo
  {pages} {451} (\bibinfo {year} {1961})}\BibitemShut {NoStop}%
\bibitem [{\citenamefont {Smerzi}\ \emph {et~al.}(1997)\citenamefont {Smerzi},
  \citenamefont {Fantoni}, \citenamefont {Giovanazzi},\ and\ \citenamefont
  {Shenoy}}]{smerzi1997}%
  \BibitemOpen
  \bibfield  {author} {\bibinfo {author} {\bibfnamefont {A.}~\bibnamefont
  {Smerzi}}, \bibinfo {author} {\bibfnamefont {S.}~\bibnamefont {Fantoni}},
  \bibinfo {author} {\bibfnamefont {S.}~\bibnamefont {Giovanazzi}}, \ and\
  \bibinfo {author} {\bibfnamefont {S.~R.}\ \bibnamefont {Shenoy}},\ }\href
  {\doibase 10.1103/PhysRevLett.79.4950} {\bibfield  {journal} {\bibinfo
  {journal} {Physical Review Letters}\ }\textbf {\bibinfo {volume} {79}},\
  \bibinfo {pages} {4950} (\bibinfo {year} {1997})}\BibitemShut {NoStop}%
\bibitem [{\citenamefont {P\'erez-Garc\'{\i}a}\ \emph
  {et~al.}(1997)\citenamefont {P\'erez-Garc\'{\i}a}, \citenamefont {Michinel},
  \citenamefont {Cirac}, \citenamefont {Lewenstein},\ and\ \citenamefont
  {Zoller}}]{perez-garcia1997}%
  \BibitemOpen
  \bibfield  {author} {\bibinfo {author} {\bibfnamefont {V.~M.}\ \bibnamefont
  {P\'erez-Garc\'{\i}a}}, \bibinfo {author} {\bibfnamefont {H.}~\bibnamefont
  {Michinel}}, \bibinfo {author} {\bibfnamefont {J.~I.}\ \bibnamefont {Cirac}},
  \bibinfo {author} {\bibfnamefont {M.}~\bibnamefont {Lewenstein}}, \ and\
  \bibinfo {author} {\bibfnamefont {P.}~\bibnamefont {Zoller}},\ }\href
  {\doibase 10.1103/PhysRevA.56.1424} {\bibfield  {journal} {\bibinfo
  {journal} {Physical Review A}\ }\textbf {\bibinfo {volume} {56}},\ \bibinfo
  {pages} {1424} (\bibinfo {year} {1997})}\BibitemShut {NoStop}%
\bibitem [{\citenamefont {Raghavan}\ \emph {et~al.}(1999)\citenamefont
  {Raghavan}, \citenamefont {Smerzi}, \citenamefont {Fantoni},\ and\
  \citenamefont {Shenoy}}]{raghavan1999}%
  \BibitemOpen
  \bibfield  {author} {\bibinfo {author} {\bibfnamefont {S.}~\bibnamefont
  {Raghavan}}, \bibinfo {author} {\bibfnamefont {A.}~\bibnamefont {Smerzi}},
  \bibinfo {author} {\bibfnamefont {S.}~\bibnamefont {Fantoni}}, \ and\
  \bibinfo {author} {\bibfnamefont {S.~R.}\ \bibnamefont {Shenoy}},\ }\href
  {\doibase 10.1103/PhysRevA.59.620} {\bibfield  {journal} {\bibinfo  {journal}
  {Physical Review A}\ }\textbf {\bibinfo {volume} {59}},\ \bibinfo {pages}
  {620} (\bibinfo {year} {1999})}\BibitemShut {NoStop}%
\bibitem [{\citenamefont {Bao}\ \emph {et~al.}(2003)\citenamefont {Bao},
  \citenamefont {Jaksch},\ and\ \citenamefont {Markowich}}]{bao2003}%
  \BibitemOpen
  \bibfield  {author} {\bibinfo {author} {\bibfnamefont {W.}~\bibnamefont
  {Bao}}, \bibinfo {author} {\bibfnamefont {D.}~\bibnamefont {Jaksch}}, \ and\
  \bibinfo {author} {\bibfnamefont {P.~A.}\ \bibnamefont {Markowich}},\ }\href
  {\doibase http://dx.doi.org/10.1016/S0021-9991(03)00102-5} {\bibfield
  {journal} {\bibinfo  {journal} {Journal of Computational Physics}\ }\textbf
  {\bibinfo {volume} {187}},\ \bibinfo {pages} {318} (\bibinfo {year}
  {2003})}\BibitemShut {NoStop}%
\bibitem [{\citenamefont {Liang}\ \emph {et~al.}(2005)\citenamefont {Liang},
  \citenamefont {Zhang},\ and\ \citenamefont {Liu}}]{liang2005}%
  \BibitemOpen
  \bibfield  {author} {\bibinfo {author} {\bibfnamefont {Z.~X.}\ \bibnamefont
  {Liang}}, \bibinfo {author} {\bibfnamefont {Z.~D.}\ \bibnamefont {Zhang}}, \
  and\ \bibinfo {author} {\bibfnamefont {W.~M.}\ \bibnamefont {Liu}},\ }\href
  {\doibase 10.1103/PhysRevLett.94.050402} {\bibfield  {journal} {\bibinfo
  {journal} {Physical Review Letters}\ }\textbf {\bibinfo {volume} {94}},\
  \bibinfo {eid} {050402} (\bibinfo {year} {2005})}\BibitemShut {NoStop}%
\bibitem [{\citenamefont {Ananikian}\ and\ \citenamefont
  {Bergeman}(2006)}]{ananikian2006}%
  \BibitemOpen
  \bibfield  {author} {\bibinfo {author} {\bibfnamefont {D.}~\bibnamefont
  {Ananikian}}\ and\ \bibinfo {author} {\bibfnamefont {T.}~\bibnamefont
  {Bergeman}},\ }\href {\doibase 10.1103/PhysRevA.73.013604} {\bibfield
  {journal} {\bibinfo  {journal} {Physical Review A}\ }\textbf {\bibinfo
  {volume} {73}},\ \bibinfo {eid} {013604} (\bibinfo {year}
  {2006})}\BibitemShut {NoStop}%
\bibitem [{\citenamefont {Leggett}(2001)}]{leggett2001}%
  \BibitemOpen
  \bibfield  {author} {\bibinfo {author} {\bibfnamefont {A.~J.}\ \bibnamefont
  {Leggett}},\ }\href {\doibase 10.1103/RevModPhys.73.307} {\bibfield
  {journal} {\bibinfo  {journal} {Reviews of Modern Physics}\ }\textbf
  {\bibinfo {volume} {73}},\ \bibinfo {pages} {307} (\bibinfo {year}
  {2001})}\BibitemShut {NoStop}%
\bibitem [{\citenamefont {Minguzzi}\ \emph {et~al.}(2004)\citenamefont
  {Minguzzi}, \citenamefont {Succi}, \citenamefont {Toschi}, \citenamefont
  {Tosi},\ and\ \citenamefont {Vignolo}}]{minguzzi2004}%
  \BibitemOpen
  \bibfield  {author} {\bibinfo {author} {\bibfnamefont {A.}~\bibnamefont
  {Minguzzi}}, \bibinfo {author} {\bibfnamefont {S.}~\bibnamefont {Succi}},
  \bibinfo {author} {\bibfnamefont {F.}~\bibnamefont {Toschi}}, \bibinfo
  {author} {\bibfnamefont {M.}~\bibnamefont {Tosi}}, \ and\ \bibinfo {author}
  {\bibfnamefont {P.}~\bibnamefont {Vignolo}},\ }\href {\doibase
  http://dx.doi.org/10.1016/j.physrep.2004.02.001} {\bibfield  {journal}
  {\bibinfo  {journal} {Physics Reports}\ }\textbf {\bibinfo {volume} {395}},\
  \bibinfo {pages} {223} (\bibinfo {year} {2004})}\BibitemShut {NoStop}%
\bibitem [{\citenamefont {Streltsov}\ \emph {et~al.}(2007)\citenamefont
  {Streltsov}, \citenamefont {Alon},\ and\ \citenamefont
  {Cederbaum}}]{streltsov2007}%
  \BibitemOpen
  \bibfield  {author} {\bibinfo {author} {\bibfnamefont {A.~I.}\ \bibnamefont
  {Streltsov}}, \bibinfo {author} {\bibfnamefont {O.~E.}\ \bibnamefont {Alon}},
  \ and\ \bibinfo {author} {\bibfnamefont {L.~S.}\ \bibnamefont {Cederbaum}},\
  }\href {\doibase 10.1103/PhysRevLett.99.030402} {\bibfield  {journal}
  {\bibinfo  {journal} {Physical Review Letters}\ }\textbf {\bibinfo {volume}
  {99}},\ \bibinfo {eid} {030402} (\bibinfo {year} {2007})}\BibitemShut
  {NoStop}%
\bibitem [{\citenamefont {Alon}\ \emph {et~al.}(2008)\citenamefont {Alon},
  \citenamefont {Streltsov},\ and\ \citenamefont {Cederbaum}}]{alon2008}%
  \BibitemOpen
  \bibfield  {author} {\bibinfo {author} {\bibfnamefont {O.~E.}\ \bibnamefont
  {Alon}}, \bibinfo {author} {\bibfnamefont {A.~I.}\ \bibnamefont {Streltsov}},
  \ and\ \bibinfo {author} {\bibfnamefont {L.~S.}\ \bibnamefont {Cederbaum}},\
  }\href {\doibase 10.1103/PhysRevA.77.033613} {\bibfield  {journal} {\bibinfo
  {journal} {Physical Review A}\ }\textbf {\bibinfo {volume} {77}},\ \bibinfo
  {eid} {033613} (\bibinfo {year} {2008})}\BibitemShut {NoStop}%
\bibitem [{\citenamefont {Streltsov}\ \emph {et~al.}(2008)\citenamefont
  {Streltsov}, \citenamefont {Alon},\ and\ \citenamefont
  {Cederbaum}}]{streltsov2008}%
  \BibitemOpen
  \bibfield  {author} {\bibinfo {author} {\bibfnamefont {A.~I.}\ \bibnamefont
  {Streltsov}}, \bibinfo {author} {\bibfnamefont {O.~E.}\ \bibnamefont {Alon}},
  \ and\ \bibinfo {author} {\bibfnamefont {L.~S.}\ \bibnamefont {Cederbaum}},\
  }\href {\doibase 10.1103/PhysRevLett.100.130401} {\bibfield  {journal}
  {\bibinfo  {journal} {Physical Review Letters}\ }\textbf {\bibinfo {volume}
  {100}},\ \bibinfo {eid} {130401} (\bibinfo {year} {2008})}\BibitemShut
  {NoStop}%
\bibitem [{\citenamefont {Streltsov}\ \emph
  {et~al.}(2009{\natexlab{a}})\citenamefont {Streltsov}, \citenamefont {Alon},\
  and\ \citenamefont {Cederbaum}}]{streltsov2009}%
  \BibitemOpen
  \bibfield  {author} {\bibinfo {author} {\bibfnamefont {A.~I.}\ \bibnamefont
  {Streltsov}}, \bibinfo {author} {\bibfnamefont {O.~E.}\ \bibnamefont {Alon}},
  \ and\ \bibinfo {author} {\bibfnamefont {L.~S.}\ \bibnamefont {Cederbaum}},\
  }\href {\doibase 10.1103/PhysRevA.80.043616} {\bibfield  {journal} {\bibinfo
  {journal} {Physical Review A}\ }\textbf {\bibinfo {volume} {80}},\ \bibinfo
  {eid} {043616} (\bibinfo {year} {2009}{\natexlab{a}})}\BibitemShut {NoStop}%
\bibitem [{\citenamefont {Streltsov}\ \emph
  {et~al.}(2009{\natexlab{b}})\citenamefont {Streltsov}, \citenamefont {Alon},\
  and\ \citenamefont {Cederbaum}}]{streltsov2009a}%
  \BibitemOpen
  \bibfield  {author} {\bibinfo {author} {\bibfnamefont {A.~I.}\ \bibnamefont
  {Streltsov}}, \bibinfo {author} {\bibfnamefont {O.~E.}\ \bibnamefont {Alon}},
  \ and\ \bibinfo {author} {\bibfnamefont {L.~S.}\ \bibnamefont {Cederbaum}},\
  }\href {http://stacks.iop.org/0953-4075/42/i=9/a=091004} {\bibfield
  {journal} {\bibinfo  {journal} {Journal of Physics B}\ }\textbf {\bibinfo
  {volume} {42}},\ \bibinfo {eid} {091004} (\bibinfo {year}
  {2009}{\natexlab{b}})}\BibitemShut {NoStop}%
\bibitem [{\citenamefont {Lode}\ \emph {et~al.}(2009)\citenamefont {Lode},
  \citenamefont {Streltsov}, \citenamefont {Alon}, \citenamefont {Meyer},\ and\
  \citenamefont {Cederbaum}}]{lode2009}%
  \BibitemOpen
  \bibfield  {author} {\bibinfo {author} {\bibfnamefont {A.~U.~J.}\
  \bibnamefont {Lode}}, \bibinfo {author} {\bibfnamefont {A.~I.}\ \bibnamefont
  {Streltsov}}, \bibinfo {author} {\bibfnamefont {O.~E.}\ \bibnamefont {Alon}},
  \bibinfo {author} {\bibfnamefont {H.-D.}\ \bibnamefont {Meyer}}, \ and\
  \bibinfo {author} {\bibfnamefont {L.~S.}\ \bibnamefont {Cederbaum}},\ }\href
  {http://stacks.iop.org/0953-4075/42/i=4/a=044018} {\bibfield  {journal}
  {\bibinfo  {journal} {Journal of Physics B}\ }\textbf {\bibinfo {volume}
  {42}},\ \bibinfo {eid} {044018} (\bibinfo {year} {2009})}\BibitemShut
  {NoStop}%
\bibitem [{\citenamefont {Sakmann}\ \emph {et~al.}(2009)\citenamefont
  {Sakmann}, \citenamefont {Streltsov}, \citenamefont {Alon},\ and\
  \citenamefont {Cederbaum}}]{sakmann2009}%
  \BibitemOpen
  \bibfield  {author} {\bibinfo {author} {\bibfnamefont {K.}~\bibnamefont
  {Sakmann}}, \bibinfo {author} {\bibfnamefont {A.~I.}\ \bibnamefont
  {Streltsov}}, \bibinfo {author} {\bibfnamefont {O.~E.}\ \bibnamefont {Alon}},
  \ and\ \bibinfo {author} {\bibfnamefont {L.~S.}\ \bibnamefont {Cederbaum}},\
  }\href {\doibase 10.1103/PhysRevLett.103.220601} {\bibfield  {journal}
  {\bibinfo  {journal} {Physical Review Letters}\ }\textbf {\bibinfo {volume}
  {103}},\ \bibinfo {eid} {220601} (\bibinfo {year} {2009})}\BibitemShut
  {NoStop}%
\bibitem [{\citenamefont {Sakmann}\ \emph {et~al.}(2010)\citenamefont
  {Sakmann}, \citenamefont {Streltsov}, \citenamefont {Alon},\ and\
  \citenamefont {Cederbaum}}]{sakmann2010}%
  \BibitemOpen
  \bibfield  {author} {\bibinfo {author} {\bibfnamefont {K.}~\bibnamefont
  {Sakmann}}, \bibinfo {author} {\bibfnamefont {A.~I.}\ \bibnamefont
  {Streltsov}}, \bibinfo {author} {\bibfnamefont {O.~E.}\ \bibnamefont {Alon}},
  \ and\ \bibinfo {author} {\bibfnamefont {L.~S.}\ \bibnamefont {Cederbaum}},\
  }\href {\doibase 10.1103/PhysRevA.82.013620} {\bibfield  {journal} {\bibinfo
  {journal} {Physical Review A}\ }\textbf {\bibinfo {volume} {82}},\ \bibinfo
  {eid} {013620} (\bibinfo {year} {2010})}\BibitemShut {NoStop}%
\bibitem [{\citenamefont {Streltsov}\ \emph
  {et~al.}(2011{\natexlab{a}})\citenamefont {Streltsov}, \citenamefont {Alon},\
  and\ \citenamefont {Cederbaum}}]{streltsov2011}%
  \BibitemOpen
  \bibfield  {author} {\bibinfo {author} {\bibfnamefont {A.~I.}\ \bibnamefont
  {Streltsov}}, \bibinfo {author} {\bibfnamefont {O.~E.}\ \bibnamefont {Alon}},
  \ and\ \bibinfo {author} {\bibfnamefont {L.~S.}\ \bibnamefont {Cederbaum}},\
  }\href {\doibase 10.1103/PhysRevLett.106.240401} {\bibfield  {journal}
  {\bibinfo  {journal} {Physical Review Letters}\ }\textbf {\bibinfo {volume}
  {106}},\ \bibinfo {eid} {240401} (\bibinfo {year}
  {2011}{\natexlab{a}})}\BibitemShut {NoStop}%
\bibitem [{\citenamefont {Streltsov}\ \emph
  {et~al.}(2011{\natexlab{b}})\citenamefont {Streltsov}, \citenamefont
  {Sakmann}, \citenamefont {Alon},\ and\ \citenamefont
  {Cederbaum}}]{streltsov2011a}%
  \BibitemOpen
  \bibfield  {author} {\bibinfo {author} {\bibfnamefont {A.~I.}\ \bibnamefont
  {Streltsov}}, \bibinfo {author} {\bibfnamefont {K.}~\bibnamefont {Sakmann}},
  \bibinfo {author} {\bibfnamefont {O.~E.}\ \bibnamefont {Alon}}, \ and\
  \bibinfo {author} {\bibfnamefont {L.~S.}\ \bibnamefont {Cederbaum}},\ }\href
  {\doibase 10.1103/PhysRevA.83.043604} {\bibfield  {journal} {\bibinfo
  {journal} {Physical Review A}\ }\textbf {\bibinfo {volume} {83}},\ \bibinfo
  {eid} {043604} (\bibinfo {year} {2011}{\natexlab{b}})}\BibitemShut {NoStop}%
\bibitem [{\citenamefont {Lode}\ \emph {et~al.}(2012)\citenamefont {Lode},
  \citenamefont {Sakmann}, \citenamefont {Alon}, \citenamefont {Cederbaum},\
  and\ \citenamefont {Streltsov}}]{lode2012}%
  \BibitemOpen
  \bibfield  {author} {\bibinfo {author} {\bibfnamefont {A.~U.~J.}\
  \bibnamefont {Lode}}, \bibinfo {author} {\bibfnamefont {K.}~\bibnamefont
  {Sakmann}}, \bibinfo {author} {\bibfnamefont {O.~E.}\ \bibnamefont {Alon}},
  \bibinfo {author} {\bibfnamefont {L.~S.}\ \bibnamefont {Cederbaum}}, \ and\
  \bibinfo {author} {\bibfnamefont {A.~I.}\ \bibnamefont {Streltsov}},\ }\href
  {\doibase 10.1103/PhysRevA.86.063606} {\bibfield  {journal} {\bibinfo
  {journal} {Physical Review A}\ }\textbf {\bibinfo {volume} {86}},\ \bibinfo
  {eid} {063606} (\bibinfo {year} {2012})}\BibitemShut {NoStop}%
\bibitem [{\citenamefont {Streltsov}(2013)}]{streltsov2013}%
  \BibitemOpen
  \bibfield  {author} {\bibinfo {author} {\bibfnamefont {A.~I.}\ \bibnamefont
  {Streltsov}},\ }\href {\doibase 10.1103/PhysRevA.88.041602} {\bibfield
  {journal} {\bibinfo  {journal} {Physical Review A}\ }\textbf {\bibinfo
  {volume} {88}},\ \bibinfo {eid} {041602(R)} (\bibinfo {year}
  {2013})}\BibitemShut {NoStop}%
\bibitem [{\citenamefont {Beinke}\ \emph {et~al.}(2015)\citenamefont {Beinke},
  \citenamefont {Klaiman}, \citenamefont {Cederbaum}, \citenamefont
  {Streltsov},\ and\ \citenamefont {Alon}}]{beinke2015}%
  \BibitemOpen
  \bibfield  {author} {\bibinfo {author} {\bibfnamefont {R.}~\bibnamefont
  {Beinke}}, \bibinfo {author} {\bibfnamefont {S.}~\bibnamefont {Klaiman}},
  \bibinfo {author} {\bibfnamefont {L.~S.}\ \bibnamefont {Cederbaum}}, \bibinfo
  {author} {\bibfnamefont {A.~I.}\ \bibnamefont {Streltsov}}, \ and\ \bibinfo
  {author} {\bibfnamefont {O.~E.}\ \bibnamefont {Alon}},\ }\href {\doibase
  10.1103/PhysRevA.92.043627} {\bibfield  {journal} {\bibinfo  {journal}
  {Physical Review A}\ }\textbf {\bibinfo {volume} {92}},\ \bibinfo {pages}
  {043627} (\bibinfo {year} {2015})}\BibitemShut {NoStop}%
\bibitem [{\citenamefont {Cao}\ \emph {et~al.}(2013)\citenamefont {Cao},
  \citenamefont {Kr\"{o}nke}, \citenamefont {Vendrell},\ and\ \citenamefont
  {Schmelcher}}]{cao2013}%
  \BibitemOpen
  \bibfield  {author} {\bibinfo {author} {\bibfnamefont {L.}~\bibnamefont
  {Cao}}, \bibinfo {author} {\bibfnamefont {S.}~\bibnamefont {Kr\"{o}nke}},
  \bibinfo {author} {\bibfnamefont {O.}~\bibnamefont {Vendrell}}, \ and\
  \bibinfo {author} {\bibfnamefont {P.}~\bibnamefont {Schmelcher}},\ }\href
  {\doibase http://dx.doi.org/10.1063/1.4821350} {\bibfield  {journal}
  {\bibinfo  {journal} {Journal of Chemical Physics}\ }\textbf {\bibinfo
  {volume} {139}},\ \bibinfo {eid} {134103} (\bibinfo {year}
  {2013})}\BibitemShut {NoStop}%
\bibitem [{\citenamefont {Kr\"{o}nke}\ \emph {et~al.}(2013)\citenamefont
  {Kr\"{o}nke}, \citenamefont {Cao}, \citenamefont {Vendrell},\ and\
  \citenamefont {Schmelcher}}]{kronke2013}%
  \BibitemOpen
  \bibfield  {author} {\bibinfo {author} {\bibfnamefont {S.}~\bibnamefont
  {Kr\"{o}nke}}, \bibinfo {author} {\bibfnamefont {L.}~\bibnamefont {Cao}},
  \bibinfo {author} {\bibfnamefont {O.}~\bibnamefont {Vendrell}}, \ and\
  \bibinfo {author} {\bibfnamefont {P.}~\bibnamefont {Schmelcher}},\ }\href
  {http://stacks.iop.org/1367-2630/15/i=6/a=063018} {\bibfield  {journal}
  {\bibinfo  {journal} {New Journal of Physics}\ }\textbf {\bibinfo {volume}
  {15}},\ \bibinfo {eid} {063018} (\bibinfo {year} {2013})}\BibitemShut
  {NoStop}%
\bibitem [{\citenamefont {Schurer}\ \emph {et~al.}(2017)\citenamefont
  {Schurer}, \citenamefont {Negretti},\ and\ \citenamefont
  {Schmelcher}}]{schurer2017}%
  \BibitemOpen
  \bibfield  {author} {\bibinfo {author} {\bibfnamefont {J.~M.}\ \bibnamefont
  {Schurer}}, \bibinfo {author} {\bibfnamefont {A.}~\bibnamefont {Negretti}}, \
  and\ \bibinfo {author} {\bibfnamefont {P.}~\bibnamefont {Schmelcher}},\
  }\href {\doibase 10.1103/PhysRevLett.119.063001} {\bibfield  {journal}
  {\bibinfo  {journal} {Physical Review Letters}\ }\textbf {\bibinfo {volume}
  {119}},\ \bibinfo {pages} {063001} (\bibinfo {year} {2017})}\BibitemShut
  {NoStop}%
\bibitem [{\citenamefont {Keiler}\ and\ \citenamefont
  {Schmelcher}(2018)}]{keiler2018}%
  \BibitemOpen
  \bibfield  {author} {\bibinfo {author} {\bibfnamefont {K.}~\bibnamefont
  {Keiler}}\ and\ \bibinfo {author} {\bibfnamefont {P.}~\bibnamefont
  {Schmelcher}},\ }\href {\doibase 10.1088/1367-2630/aae98f} {\bibfield
  {journal} {\bibinfo  {journal} {New Journal of Physics}\ }\textbf {\bibinfo
  {volume} {20}},\ \bibinfo {pages} {103042} (\bibinfo {year}
  {2018})}\BibitemShut {NoStop}%
\bibitem [{\citenamefont {Mistakidis}\ \emph
  {et~al.}(2018{\natexlab{a}})\citenamefont {Mistakidis}, \citenamefont
  {Volosniev}, \citenamefont {Zinner},\ and\ \citenamefont
  {Schmelcher}}]{mistakidis2018b}%
  \BibitemOpen
  \bibfield  {author} {\bibinfo {author} {\bibfnamefont {S.~I.}\ \bibnamefont
  {Mistakidis}}, \bibinfo {author} {\bibfnamefont {A.~G.}\ \bibnamefont
  {Volosniev}}, \bibinfo {author} {\bibfnamefont {N.~T.}\ \bibnamefont
  {Zinner}}, \ and\ \bibinfo {author} {\bibfnamefont {P.}~\bibnamefont
  {Schmelcher}},\ }\href {https://arxiv.org/abs/1809.01889} {\bibfield
  {journal} {\bibinfo  {journal} {ArXiv e-prints}\ } (\bibinfo {year}
  {2018}{\natexlab{a}})},\ \bibinfo {note} {arXiv:1809.01889}\BibitemShut
  {NoStop}%
\bibitem [{\citenamefont {Mistakidis}\ \emph
  {et~al.}(2018{\natexlab{b}})\citenamefont {Mistakidis}, \citenamefont
  {Katsimiga}, \citenamefont {Koutentakis}, \citenamefont {Busch},\ and\
  \citenamefont {Schmelcher}}]{mistakidis2018c}%
  \BibitemOpen
  \bibfield  {author} {\bibinfo {author} {\bibfnamefont {S.}~\bibnamefont
  {Mistakidis}}, \bibinfo {author} {\bibfnamefont {G.}~\bibnamefont
  {Katsimiga}}, \bibinfo {author} {\bibfnamefont {G.}~\bibnamefont
  {Koutentakis}}, \bibinfo {author} {\bibfnamefont {T.}~\bibnamefont {Busch}},
  \ and\ \bibinfo {author} {\bibfnamefont {P.}~\bibnamefont {Schmelcher}},\
  }\href {https://arxiv.org/abs/1811.10702} {\bibfield  {journal} {\bibinfo
  {journal} {ArXiv e-prints}\ } (\bibinfo {year} {2018}{\natexlab{b}})},\
  \bibinfo {note} {arXiv:1811.10702}\BibitemShut {NoStop}%
\bibitem [{\citenamefont {Katsimiga}\ \emph {et~al.}(2018)\citenamefont
  {Katsimiga}, \citenamefont {Mistakidis}, \citenamefont {Koutentakis},
  \citenamefont {Kevrekidis},\ and\ \citenamefont
  {Schmelcher}}]{katsimiga2018}%
  \BibitemOpen
  \bibfield  {author} {\bibinfo {author} {\bibfnamefont {G.~C.}\ \bibnamefont
  {Katsimiga}}, \bibinfo {author} {\bibfnamefont {S.~I.}\ \bibnamefont
  {Mistakidis}}, \bibinfo {author} {\bibfnamefont {G.~M.}\ \bibnamefont
  {Koutentakis}}, \bibinfo {author} {\bibfnamefont {P.~G.}\ \bibnamefont
  {Kevrekidis}}, \ and\ \bibinfo {author} {\bibfnamefont {P.}~\bibnamefont
  {Schmelcher}},\ }\href {\doibase 10.1103/PhysRevA.98.013632} {\bibfield
  {journal} {\bibinfo  {journal} {Physical Review A}\ }\textbf {\bibinfo
  {volume} {98}},\ \bibinfo {pages} {013632} (\bibinfo {year}
  {2018})}\BibitemShut {NoStop}%
\bibitem [{\citenamefont {Mistakidis}\ \emph
  {et~al.}(2018{\natexlab{c}})\citenamefont {Mistakidis}, \citenamefont
  {Katsimiga}, \citenamefont {Kevrekidis},\ and\ \citenamefont
  {Schmelcher}}]{mistakidis2018}%
  \BibitemOpen
  \bibfield  {author} {\bibinfo {author} {\bibfnamefont {S.~I.}\ \bibnamefont
  {Mistakidis}}, \bibinfo {author} {\bibfnamefont {G.~C.}\ \bibnamefont
  {Katsimiga}}, \bibinfo {author} {\bibfnamefont {P.~G.}\ \bibnamefont
  {Kevrekidis}}, \ and\ \bibinfo {author} {\bibfnamefont {P.}~\bibnamefont
  {Schmelcher}},\ }\href {\doibase 10.1088/1367-2630/aabc6a} {\bibfield
  {journal} {\bibinfo  {journal} {New Journal of Physics}\ }\textbf {\bibinfo
  {volume} {20}},\ \bibinfo {pages} {043052} (\bibinfo {year}
  {2018}{\natexlab{c}})}\BibitemShut {NoStop}%
\bibitem [{\citenamefont {Katsimiga}\ \emph
  {et~al.}(2017{\natexlab{a}})\citenamefont {Katsimiga}, \citenamefont
  {Koutentakis}, \citenamefont {Mistakidis}, \citenamefont {Kevrekidis},\ and\
  \citenamefont {Schmelcher}}]{katsimiga2017}%
  \BibitemOpen
  \bibfield  {author} {\bibinfo {author} {\bibfnamefont {G.~C.}\ \bibnamefont
  {Katsimiga}}, \bibinfo {author} {\bibfnamefont {G.~M.}\ \bibnamefont
  {Koutentakis}}, \bibinfo {author} {\bibfnamefont {S.~I.}\ \bibnamefont
  {Mistakidis}}, \bibinfo {author} {\bibfnamefont {P.~G.}\ \bibnamefont
  {Kevrekidis}}, \ and\ \bibinfo {author} {\bibfnamefont {P.}~\bibnamefont
  {Schmelcher}},\ }\href {\doibase 10.1088/1367-2630/aa766b} {\bibfield
  {journal} {\bibinfo  {journal} {New Journal of Physics}\ }\textbf {\bibinfo
  {volume} {19}},\ \bibinfo {pages} {073004} (\bibinfo {year}
  {2017}{\natexlab{a}})}\BibitemShut {NoStop}%
\bibitem [{\citenamefont {Kr\"onke}\ and\ \citenamefont
  {Schmelcher}(2015)}]{kronke2015}%
  \BibitemOpen
  \bibfield  {author} {\bibinfo {author} {\bibfnamefont {S.}~\bibnamefont
  {Kr\"onke}}\ and\ \bibinfo {author} {\bibfnamefont {P.}~\bibnamefont
  {Schmelcher}},\ }\href {\doibase 10.1103/PhysRevA.91.053614} {\bibfield
  {journal} {\bibinfo  {journal} {Physical Review A}\ }\textbf {\bibinfo
  {volume} {91}},\ \bibinfo {pages} {053614} (\bibinfo {year}
  {2015})}\BibitemShut {NoStop}%
\bibitem [{\citenamefont {Katsimiga}\ \emph
  {et~al.}(2017{\natexlab{b}})\citenamefont {Katsimiga}, \citenamefont
  {Mistakidis}, \citenamefont {Koutentakis}, \citenamefont {Kevrekidis},\ and\
  \citenamefont {Schmelcher}}]{katsimiga2017a}%
  \BibitemOpen
  \bibfield  {author} {\bibinfo {author} {\bibfnamefont {G.~C.}\ \bibnamefont
  {Katsimiga}}, \bibinfo {author} {\bibfnamefont {S.~I.}\ \bibnamefont
  {Mistakidis}}, \bibinfo {author} {\bibfnamefont {G.~M.}\ \bibnamefont
  {Koutentakis}}, \bibinfo {author} {\bibfnamefont {P.~G.}\ \bibnamefont
  {Kevrekidis}}, \ and\ \bibinfo {author} {\bibfnamefont {P.}~\bibnamefont
  {Schmelcher}},\ }\href {\doibase 10.1088/1367-2630/aa96f6} {\bibfield
  {journal} {\bibinfo  {journal} {New Journal of Physics}\ }\textbf {\bibinfo
  {volume} {19}},\ \bibinfo {pages} {123012} (\bibinfo {year}
  {2017}{\natexlab{b}})}\BibitemShut {NoStop}%
\bibitem [{\citenamefont {Mistakidis}\ \emph {et~al.}(2014)\citenamefont
  {Mistakidis}, \citenamefont {Cao},\ and\ \citenamefont
  {Schmelcher}}]{mistakidis2014}%
  \BibitemOpen
  \bibfield  {author} {\bibinfo {author} {\bibfnamefont {S.~I.}\ \bibnamefont
  {Mistakidis}}, \bibinfo {author} {\bibfnamefont {L.}~\bibnamefont {Cao}}, \
  and\ \bibinfo {author} {\bibfnamefont {P.}~\bibnamefont {Schmelcher}},\
  }\href {\doibase 10.1088/0953-4075/47/22/225303} {\bibfield  {journal}
  {\bibinfo  {journal} {Journal of Physics B: Atomic, Molecular and Optical
  Physics}\ }\textbf {\bibinfo {volume} {47}},\ \bibinfo {pages} {225303}
  (\bibinfo {year} {2014})}\BibitemShut {NoStop}%
\bibitem [{\citenamefont {Mistakidis}\ \emph
  {et~al.}(2015{\natexlab{a}})\citenamefont {Mistakidis}, \citenamefont {Cao},\
  and\ \citenamefont {Schmelcher}}]{mistakidis2015}%
  \BibitemOpen
  \bibfield  {author} {\bibinfo {author} {\bibfnamefont {S.~I.}\ \bibnamefont
  {Mistakidis}}, \bibinfo {author} {\bibfnamefont {L.}~\bibnamefont {Cao}}, \
  and\ \bibinfo {author} {\bibfnamefont {P.}~\bibnamefont {Schmelcher}},\
  }\href {\doibase 10.1103/PhysRevA.91.033611} {\bibfield  {journal} {\bibinfo
  {journal} {Physical Review A}\ }\textbf {\bibinfo {volume} {91}},\ \bibinfo
  {pages} {033611} (\bibinfo {year} {2015}{\natexlab{a}})}\BibitemShut
  {NoStop}%
\bibitem [{\citenamefont {Mistakidis}\ \emph
  {et~al.}(2015{\natexlab{b}})\citenamefont {Mistakidis}, \citenamefont {Wulf},
  \citenamefont {Negretti},\ and\ \citenamefont
  {Schmelcher}}]{mistakidis2015a}%
  \BibitemOpen
  \bibfield  {author} {\bibinfo {author} {\bibfnamefont {S.~I.}\ \bibnamefont
  {Mistakidis}}, \bibinfo {author} {\bibfnamefont {T.}~\bibnamefont {Wulf}},
  \bibinfo {author} {\bibfnamefont {A.}~\bibnamefont {Negretti}}, \ and\
  \bibinfo {author} {\bibfnamefont {P.}~\bibnamefont {Schmelcher}},\ }\href
  {\doibase 10.1088/0953-4075/48/24/244004} {\bibfield  {journal} {\bibinfo
  {journal} {Journal of Physics B: Atomic, Molecular and Optical Physics}\
  }\textbf {\bibinfo {volume} {48}},\ \bibinfo {pages} {244004} (\bibinfo
  {year} {2015}{\natexlab{b}})}\BibitemShut {NoStop}%
\bibitem [{\citenamefont {Mistakidis}\ and\ \citenamefont
  {Schmelcher}(2017)}]{mistakidis2017b}%
  \BibitemOpen
  \bibfield  {author} {\bibinfo {author} {\bibfnamefont {S.~I.}\ \bibnamefont
  {Mistakidis}}\ and\ \bibinfo {author} {\bibfnamefont {P.}~\bibnamefont
  {Schmelcher}},\ }\href {\doibase 10.1103/PhysRevA.95.013625} {\bibfield
  {journal} {\bibinfo  {journal} {Physical Review A}\ }\textbf {\bibinfo
  {volume} {95}},\ \bibinfo {pages} {013625} (\bibinfo {year}
  {2017})}\BibitemShut {NoStop}%
\bibitem [{\citenamefont {Koutentakis}\ \emph {et~al.}(2017)\citenamefont
  {Koutentakis}, \citenamefont {Mistakidis},\ and\ \citenamefont
  {Schmelcher}}]{koutentakis2017}%
  \BibitemOpen
  \bibfield  {author} {\bibinfo {author} {\bibfnamefont {G.~M.}\ \bibnamefont
  {Koutentakis}}, \bibinfo {author} {\bibfnamefont {S.~I.}\ \bibnamefont
  {Mistakidis}}, \ and\ \bibinfo {author} {\bibfnamefont {P.}~\bibnamefont
  {Schmelcher}},\ }\href {\doibase 10.1103/PhysRevA.95.013617} {\bibfield
  {journal} {\bibinfo  {journal} {Physical Review A}\ }\textbf {\bibinfo
  {volume} {95}},\ \bibinfo {pages} {013617} (\bibinfo {year}
  {2017})}\BibitemShut {NoStop}%
\bibitem [{\citenamefont {Neuhaus-Steinmetz}\ \emph {et~al.}(2017)\citenamefont
  {Neuhaus-Steinmetz}, \citenamefont {Mistakidis},\ and\ \citenamefont
  {Schmelcher}}]{neuhaus-steinmetz2017}%
  \BibitemOpen
  \bibfield  {author} {\bibinfo {author} {\bibfnamefont {J.}~\bibnamefont
  {Neuhaus-Steinmetz}}, \bibinfo {author} {\bibfnamefont {S.~I.}\ \bibnamefont
  {Mistakidis}}, \ and\ \bibinfo {author} {\bibfnamefont {P.}~\bibnamefont
  {Schmelcher}},\ }\href {\doibase 10.1103/PhysRevA.95.053610} {\bibfield
  {journal} {\bibinfo  {journal} {Physical Review A}\ }\textbf {\bibinfo
  {volume} {95}},\ \bibinfo {pages} {053610} (\bibinfo {year}
  {2017})}\BibitemShut {NoStop}%
\bibitem [{\citenamefont {Mistakidis}\ \emph
  {et~al.}(2018{\natexlab{d}})\citenamefont {Mistakidis}, \citenamefont
  {Koutentakis},\ and\ \citenamefont {Schmelcher}}]{mistakidis2018a}%
  \BibitemOpen
  \bibfield  {author} {\bibinfo {author} {\bibfnamefont {S.}~\bibnamefont
  {Mistakidis}}, \bibinfo {author} {\bibfnamefont {G.}~\bibnamefont
  {Koutentakis}}, \ and\ \bibinfo {author} {\bibfnamefont {P.}~\bibnamefont
  {Schmelcher}},\ }\href {\doibase
  https://doi.org/10.1016/j.chemphys.2017.11.022} {\bibfield  {journal}
  {\bibinfo  {journal} {Chemical Physics}\ }\textbf {\bibinfo {volume} {509}},\
  \bibinfo {pages} {106 } (\bibinfo {year} {2018}{\natexlab{d}})}\BibitemShut
  {NoStop}%
\bibitem [{\citenamefont {Pla{\ss}mann}\ \emph {et~al.}(2018)\citenamefont
  {Pla{\ss}mann}, \citenamefont {Mistakidis},\ and\ \citenamefont
  {Schmelcher}}]{plassmann2018}%
  \BibitemOpen
  \bibfield  {author} {\bibinfo {author} {\bibfnamefont {T.}~\bibnamefont
  {Pla{\ss}mann}}, \bibinfo {author} {\bibfnamefont {S.~I.}\ \bibnamefont
  {Mistakidis}}, \ and\ \bibinfo {author} {\bibfnamefont {P.}~\bibnamefont
  {Schmelcher}},\ }\href {\doibase 10.1088/1361-6455/aae57a} {\bibfield
  {journal} {\bibinfo  {journal} {Journal of Physics B: Atomic, Molecular and
  Optical Physics}\ }\textbf {\bibinfo {volume} {51}},\ \bibinfo {pages}
  {225001} (\bibinfo {year} {2018})}\BibitemShut {NoStop}%
\bibitem [{\citenamefont {Meyer}\ \emph {et~al.}(1990)\citenamefont {Meyer},
  \citenamefont {Manthe},\ and\ \citenamefont {Cederbaum}}]{meyer1990}%
  \BibitemOpen
  \bibfield  {author} {\bibinfo {author} {\bibfnamefont {H.-D.}\ \bibnamefont
  {Meyer}}, \bibinfo {author} {\bibfnamefont {U.}~\bibnamefont {Manthe}}, \
  and\ \bibinfo {author} {\bibfnamefont {L.}~\bibnamefont {Cederbaum}},\ }\href
  {\doibase 10.1016/0009-2614(90)87014-I} {\bibfield  {journal} {\bibinfo
  {journal} {Chemical Physics Letters}\ }\textbf {\bibinfo {volume} {165}},\
  \bibinfo {pages} {73} (\bibinfo {year} {1990})}\BibitemShut {NoStop}%
\bibitem [{\citenamefont {Wang}\ and\ \citenamefont {Thoss}(2003)}]{wang2003}%
  \BibitemOpen
  \bibfield  {author} {\bibinfo {author} {\bibfnamefont {H.}~\bibnamefont
  {Wang}}\ and\ \bibinfo {author} {\bibfnamefont {M.}~\bibnamefont {Thoss}},\
  }\href {\doibase 10.1063/1.1580111} {\bibfield  {journal} {\bibinfo
  {journal} {Journal of Chemical Physics}\ }\textbf {\bibinfo {volume} {119}},\
  \bibinfo {pages} {1289} (\bibinfo {year} {2003})}\BibitemShut {NoStop}%
\bibitem [{\citenamefont {Manthe}(2008)}]{manthe2008}%
  \BibitemOpen
  \bibfield  {author} {\bibinfo {author} {\bibfnamefont {U.}~\bibnamefont
  {Manthe}},\ }\href {\doibase http://dx.doi.org/10.1063/1.2902982} {\bibfield
  {journal} {\bibinfo  {journal} {Journal of Chemical Physics}\ }\textbf
  {\bibinfo {volume} {128}},\ \bibinfo {eid} {164116} (\bibinfo {year}
  {2008})}\BibitemShut {NoStop}%
\bibitem [{\citenamefont {Shalashilin}\ and\ \citenamefont
  {Child}(2000)}]{shalashilin2000}%
  \BibitemOpen
  \bibfield  {author} {\bibinfo {author} {\bibfnamefont {D.~V.}\ \bibnamefont
  {Shalashilin}}\ and\ \bibinfo {author} {\bibfnamefont {M.~S.}\ \bibnamefont
  {Child}},\ }\href {\doibase 10.1063/1.1322075} {\bibfield  {journal}
  {\bibinfo  {journal} {Journal of Chemical Physics}\ }\textbf {\bibinfo
  {volume} {113}},\ \bibinfo {pages} {10028} (\bibinfo {year}
  {2000})}\BibitemShut {NoStop}%
\bibitem [{\citenamefont {Shalashilin}\ and\ \citenamefont
  {Child}(2004{\natexlab{a}})}]{shalashilin2004}%
  \BibitemOpen
  \bibfield  {author} {\bibinfo {author} {\bibfnamefont {D.~V.}\ \bibnamefont
  {Shalashilin}}\ and\ \bibinfo {author} {\bibfnamefont {M.~S.}\ \bibnamefont
  {Child}},\ }\href {\doibase 10.1016/j.chemphys.2004.06.013} {\bibfield
  {journal} {\bibinfo  {journal} {Chemical Physics}\ }\textbf {\bibinfo
  {volume} {304}},\ \bibinfo {pages} {103} (\bibinfo {year}
  {2004}{\natexlab{a}})}\BibitemShut {NoStop}%
\bibitem [{\citenamefont {Shalashilin}\ and\ \citenamefont
  {Child}(2003)}]{shalashilin2003}%
  \BibitemOpen
  \bibfield  {author} {\bibinfo {author} {\bibfnamefont {D.~V.}\ \bibnamefont
  {Shalashilin}}\ and\ \bibinfo {author} {\bibfnamefont {M.~S.}\ \bibnamefont
  {Child}},\ }\href {\doibase 10.1063/1.1584663} {\bibfield  {journal}
  {\bibinfo  {journal} {Journal of Chemical Physics}\ }\textbf {\bibinfo
  {volume} {119}},\ \bibinfo {pages} {1961} (\bibinfo {year}
  {2003})}\BibitemShut {NoStop}%
\bibitem [{\citenamefont {Shalashilin}\ and\ \citenamefont
  {Child}(2008)}]{shalashilin2008}%
  \BibitemOpen
  \bibfield  {author} {\bibinfo {author} {\bibfnamefont {D.~V.}\ \bibnamefont
  {Shalashilin}}\ and\ \bibinfo {author} {\bibfnamefont {M.~S.}\ \bibnamefont
  {Child}},\ }\href {\doibase 10.1063/1.2828509} {\bibfield  {journal}
  {\bibinfo  {journal} {Journal of Chemical Physics}\ }\textbf {\bibinfo
  {volume} {128}},\ \bibinfo {eid} {054102} (\bibinfo {year}
  {2008})}\BibitemShut {NoStop}%
\bibitem [{\citenamefont {Simon}\ and\ \citenamefont
  {Strunz}(2014)}]{simon2014}%
  \BibitemOpen
  \bibfield  {author} {\bibinfo {author} {\bibfnamefont {L.}~\bibnamefont
  {Simon}}\ and\ \bibinfo {author} {\bibfnamefont {W.~T.}\ \bibnamefont
  {Strunz}},\ }\href {\doibase 10.1103/PhysRevA.89.052112} {\bibfield
  {journal} {\bibinfo  {journal} {Physical Review A}\ }\textbf {\bibinfo
  {volume} {89}},\ \bibinfo {pages} {052112} (\bibinfo {year}
  {2014})}\BibitemShut {NoStop}%
\bibitem [{\citenamefont {Ray}\ \emph {et~al.}(2016)\citenamefont {Ray},
  \citenamefont {Ostmann}, \citenamefont {Simon}, \citenamefont {Grossmann},\
  and\ \citenamefont {Strunz}}]{ray2016}%
  \BibitemOpen
  \bibfield  {author} {\bibinfo {author} {\bibfnamefont {S.}~\bibnamefont
  {Ray}}, \bibinfo {author} {\bibfnamefont {P.}~\bibnamefont {Ostmann}},
  \bibinfo {author} {\bibfnamefont {L.}~\bibnamefont {Simon}}, \bibinfo
  {author} {\bibfnamefont {F.}~\bibnamefont {Grossmann}}, \ and\ \bibinfo
  {author} {\bibfnamefont {W.~T.}\ \bibnamefont {Strunz}},\ }\href
  {http://stacks.iop.org/1751-8121/49/i=16/a=165303} {\bibfield  {journal}
  {\bibinfo  {journal} {Journal of Physics A: Mathematical and Theoretical}\
  }\textbf {\bibinfo {volume} {49}},\ \bibinfo {pages} {165303} (\bibinfo
  {year} {2016})}\BibitemShut {NoStop}%
\bibitem [{\citenamefont {Child}\ and\ \citenamefont
  {Shalashilin}(2003)}]{child2003}%
  \BibitemOpen
  \bibfield  {author} {\bibinfo {author} {\bibfnamefont {M.~S.}\ \bibnamefont
  {Child}}\ and\ \bibinfo {author} {\bibfnamefont {D.~V.}\ \bibnamefont
  {Shalashilin}},\ }\href {\doibase 10.1063/1.1531997} {\bibfield  {journal}
  {\bibinfo  {journal} {Journal of Chemical Physics}\ }\textbf {\bibinfo
  {volume} {118}},\ \bibinfo {pages} {2061} (\bibinfo {year}
  {2003})}\BibitemShut {NoStop}%
\bibitem [{\citenamefont {Shalashilin}\ and\ \citenamefont
  {Child}(2004{\natexlab{b}})}]{shalashilin2004a}%
  \BibitemOpen
  \bibfield  {author} {\bibinfo {author} {\bibfnamefont {D.~V.}\ \bibnamefont
  {Shalashilin}}\ and\ \bibinfo {author} {\bibfnamefont {M.~S.}\ \bibnamefont
  {Child}},\ }\href {\doibase 10.1063/1.1776111} {\bibfield  {journal}
  {\bibinfo  {journal} {Journal of Chemical Physics}\ }\textbf {\bibinfo
  {volume} {121}},\ \bibinfo {pages} {3563} (\bibinfo {year}
  {2004}{\natexlab{b}})}\BibitemShut {NoStop}%
\bibitem [{\citenamefont {Modugno}\ \emph {et~al.}(2002)\citenamefont
  {Modugno}, \citenamefont {Modugno}, \citenamefont {Riboli}, \citenamefont
  {Roati},\ and\ \citenamefont {Inguscio}}]{modugno2002}%
  \BibitemOpen
  \bibfield  {author} {\bibinfo {author} {\bibfnamefont {G.}~\bibnamefont
  {Modugno}}, \bibinfo {author} {\bibfnamefont {M.}~\bibnamefont {Modugno}},
  \bibinfo {author} {\bibfnamefont {F.}~\bibnamefont {Riboli}}, \bibinfo
  {author} {\bibfnamefont {G.}~\bibnamefont {Roati}}, \ and\ \bibinfo {author}
  {\bibfnamefont {M.}~\bibnamefont {Inguscio}},\ }\href {\doibase
  10.1103/PhysRevLett.89.190404} {\bibfield  {journal} {\bibinfo  {journal}
  {Physical Review Letters}\ }\textbf {\bibinfo {volume} {89}},\ \bibinfo
  {pages} {190404} (\bibinfo {year} {2002})}\BibitemShut {NoStop}%
\bibitem [{\citenamefont {Kawaguchi}\ and\ \citenamefont
  {Ueda}(2012)}]{kawaguchi2012}%
  \BibitemOpen
  \bibfield  {author} {\bibinfo {author} {\bibfnamefont {Y.}~\bibnamefont
  {Kawaguchi}}\ and\ \bibinfo {author} {\bibfnamefont {M.}~\bibnamefont
  {Ueda}},\ }\href {\doibase https://doi.org/10.1016/j.physrep.2012.07.005}
  {\bibfield  {journal} {\bibinfo  {journal} {Physics Reports}\ }\textbf
  {\bibinfo {volume} {520}},\ \bibinfo {pages} {253 } (\bibinfo {year}
  {2012})}\BibitemShut {NoStop}%
\bibitem [{\citenamefont {Becker}\ \emph {et~al.}(2008)\citenamefont {Becker},
  \citenamefont {Stellmer}, \citenamefont {Soltan-Panahi}, \citenamefont
  {D{\"o}rscher}, \citenamefont {Baumert}, \citenamefont {Richter},
  \citenamefont {Kronj{\"a}ger}, \citenamefont {Bongs},\ and\ \citenamefont
  {Sengstock}}]{becker2008}%
  \BibitemOpen
  \bibfield  {author} {\bibinfo {author} {\bibfnamefont {C.}~\bibnamefont
  {Becker}}, \bibinfo {author} {\bibfnamefont {S.}~\bibnamefont {Stellmer}},
  \bibinfo {author} {\bibfnamefont {P.}~\bibnamefont {Soltan-Panahi}}, \bibinfo
  {author} {\bibfnamefont {S.}~\bibnamefont {D{\"o}rscher}}, \bibinfo {author}
  {\bibfnamefont {M.}~\bibnamefont {Baumert}}, \bibinfo {author} {\bibfnamefont
  {E.-M.}\ \bibnamefont {Richter}}, \bibinfo {author} {\bibfnamefont
  {J.}~\bibnamefont {Kronj{\"a}ger}}, \bibinfo {author} {\bibfnamefont
  {K.}~\bibnamefont {Bongs}}, \ and\ \bibinfo {author} {\bibfnamefont
  {K.}~\bibnamefont {Sengstock}},\ }\href {https://doi.org/10.1038/nphys962}
  {\bibfield  {journal} {\bibinfo  {journal} {Nature Physics}\ }\textbf
  {\bibinfo {volume} {4}},\ \bibinfo {pages} {496} (\bibinfo {year}
  {2008})}\BibitemShut {NoStop}%
\bibitem [{\citenamefont {Bruderer}\ \emph {et~al.}(2007)\citenamefont
  {Bruderer}, \citenamefont {Klein}, \citenamefont {Clark},\ and\ \citenamefont
  {Jaksch}}]{bruderer2007}%
  \BibitemOpen
  \bibfield  {author} {\bibinfo {author} {\bibfnamefont {M.}~\bibnamefont
  {Bruderer}}, \bibinfo {author} {\bibfnamefont {A.}~\bibnamefont {Klein}},
  \bibinfo {author} {\bibfnamefont {S.~R.}\ \bibnamefont {Clark}}, \ and\
  \bibinfo {author} {\bibfnamefont {D.}~\bibnamefont {Jaksch}},\ }\href
  {\doibase 10.1103/PhysRevA.76.011605} {\bibfield  {journal} {\bibinfo
  {journal} {Physical Review A}\ }\textbf {\bibinfo {volume} {76}},\ \bibinfo
  {pages} {011605(R)} (\bibinfo {year} {2007})}\BibitemShut {NoStop}%
\bibitem [{\citenamefont {Greiner}\ \emph {et~al.}(2002)\citenamefont
  {Greiner}, \citenamefont {Mandel}, \citenamefont {Esslinger}, \citenamefont
  {H{\"a}nsch},\ and\ \citenamefont {Bloch}}]{greiner2002}%
  \BibitemOpen
  \bibfield  {author} {\bibinfo {author} {\bibfnamefont {M.}~\bibnamefont
  {Greiner}}, \bibinfo {author} {\bibfnamefont {O.}~\bibnamefont {Mandel}},
  \bibinfo {author} {\bibfnamefont {T.}~\bibnamefont {Esslinger}}, \bibinfo
  {author} {\bibfnamefont {T.~W.}\ \bibnamefont {H{\"a}nsch}}, \ and\ \bibinfo
  {author} {\bibfnamefont {I.}~\bibnamefont {Bloch}},\ }\href {\doibase
  10.1038/415039a} {\bibfield  {journal} {\bibinfo  {journal} {Nature}\
  }\textbf {\bibinfo {volume} {415}},\ \bibinfo {pages} {39} (\bibinfo {year}
  {2002})}\BibitemShut {NoStop}%
\bibitem [{\citenamefont {Jaksch}\ and\ \citenamefont
  {Zoller}(2005)}]{jaksch2005}%
  \BibitemOpen
  \bibfield  {author} {\bibinfo {author} {\bibfnamefont {D.}~\bibnamefont
  {Jaksch}}\ and\ \bibinfo {author} {\bibfnamefont {P.}~\bibnamefont
  {Zoller}},\ }\href {\doibase https://doi.org/10.1016/j.aop.2004.09.010}
  {\bibfield  {journal} {\bibinfo  {journal} {Annals of Physics}\ }\textbf
  {\bibinfo {volume} {315}},\ \bibinfo {pages} {52 } (\bibinfo {year}
  {2005})}\BibitemShut {NoStop}%
\bibitem [{\citenamefont {Gati}\ \emph {et~al.}(2006)\citenamefont {Gati},
  \citenamefont {Albiez}, \citenamefont {F{\"o}lling}, \citenamefont
  {Hemmerling},\ and\ \citenamefont {Oberthaler}}]{gati2006}%
  \BibitemOpen
  \bibfield  {author} {\bibinfo {author} {\bibfnamefont {R.}~\bibnamefont
  {Gati}}, \bibinfo {author} {\bibfnamefont {M.}~\bibnamefont {Albiez}},
  \bibinfo {author} {\bibfnamefont {J.}~\bibnamefont {F{\"o}lling}}, \bibinfo
  {author} {\bibfnamefont {B.}~\bibnamefont {Hemmerling}}, \ and\ \bibinfo
  {author} {\bibfnamefont {M.}~\bibnamefont {Oberthaler}},\ }\href {\doibase
  10.1007/s00340-005-2059-z} {\bibfield  {journal} {\bibinfo  {journal}
  {Applied Physics B}\ }\textbf {\bibinfo {volume} {82}},\ \bibinfo {pages}
  {207} (\bibinfo {year} {2006})}\BibitemShut {NoStop}%
\bibitem [{\citenamefont {Gati}\ and\ \citenamefont
  {Oberthaler}(2007)}]{gati2007}%
  \BibitemOpen
  \bibfield  {author} {\bibinfo {author} {\bibfnamefont {R.}~\bibnamefont
  {Gati}}\ and\ \bibinfo {author} {\bibfnamefont {M.~K.}\ \bibnamefont
  {Oberthaler}},\ }\href {\doibase 10.1088/0953-4075/40/10/r01} {\bibfield
  {journal} {\bibinfo  {journal} {Journal of Physics B: Atomic, Molecular and
  Optical Physics}\ }\textbf {\bibinfo {volume} {40}},\ \bibinfo {pages} {R61}
  (\bibinfo {year} {2007})}\BibitemShut {NoStop}%
\bibitem [{\citenamefont {Wu}\ and\ \citenamefont {Batista}(2004)}]{wu2004}%
  \BibitemOpen
  \bibfield  {author} {\bibinfo {author} {\bibfnamefont {Y.}~\bibnamefont
  {Wu}}\ and\ \bibinfo {author} {\bibfnamefont {V.~S.}\ \bibnamefont
  {Batista}},\ }\href {\doibase http://dx.doi.org/10.1063/1.1766298} {\bibfield
   {journal} {\bibinfo  {journal} {Journal of Chemical Physics}\ }\textbf
  {\bibinfo {volume} {121}},\ \bibinfo {pages} {1676} (\bibinfo {year}
  {2004})}\BibitemShut {NoStop}%
\bibitem [{\citenamefont {Sherratt}\ \emph {et~al.}(2006)\citenamefont
  {Sherratt}, \citenamefont {Shalashilin},\ and\ \citenamefont
  {Child}}]{sherratt2006}%
  \BibitemOpen
  \bibfield  {author} {\bibinfo {author} {\bibfnamefont {P.~A.~J.}\
  \bibnamefont {Sherratt}}, \bibinfo {author} {\bibfnamefont {D.~V.}\
  \bibnamefont {Shalashilin}}, \ and\ \bibinfo {author} {\bibfnamefont {M.~S.}\
  \bibnamefont {Child}},\ }\href {\doibase 10.1016/j.chemphys.2005.06.050}
  {\bibfield  {journal} {\bibinfo  {journal} {Chemical Physics}\ }\textbf
  {\bibinfo {volume} {322}},\ \bibinfo {pages} {127} (\bibinfo {year}
  {2006})}\BibitemShut {NoStop}%
\bibitem [{\citenamefont {Habershon}(2012)}]{habershon2012}%
  \BibitemOpen
  \bibfield  {author} {\bibinfo {author} {\bibfnamefont {S.}~\bibnamefont
  {Habershon}},\ }\href {\doibase http://dx.doi.org/10.1063/1.3681167}
  {\bibfield  {journal} {\bibinfo  {journal} {Journal of Chemical Physics}\
  }\textbf {\bibinfo {volume} {136}},\ \bibinfo {eid} {054109} (\bibinfo {year}
  {2012})}\BibitemShut {NoStop}%
\bibitem [{\citenamefont {Saller}\ and\ \citenamefont
  {Habershon}(2017)}]{saller2017}%
  \BibitemOpen
  \bibfield  {author} {\bibinfo {author} {\bibfnamefont {M.~A.~C.}\
  \bibnamefont {Saller}}\ and\ \bibinfo {author} {\bibfnamefont
  {S.}~\bibnamefont {Habershon}},\ }\href {\doibase 10.1021/acs.jctc.7b00021}
  {\bibfield  {journal} {\bibinfo  {journal} {Journal of Chemical Theory and
  Computation}\ }\textbf {\bibinfo {volume} {13}},\ \bibinfo {pages} {3085}
  (\bibinfo {year} {2017})}\BibitemShut {NoStop}%
\bibitem [{\citenamefont {Alborzpour}\ \emph {et~al.}(2016)\citenamefont
  {Alborzpour}, \citenamefont {Tew},\ and\ \citenamefont
  {Habershon}}]{alborzpour2016}%
  \BibitemOpen
  \bibfield  {author} {\bibinfo {author} {\bibfnamefont {J.~P.}\ \bibnamefont
  {Alborzpour}}, \bibinfo {author} {\bibfnamefont {D.~P.}\ \bibnamefont {Tew}},
  \ and\ \bibinfo {author} {\bibfnamefont {S.}~\bibnamefont {Habershon}},\
  }\href {\doibase 10.1063/1.4964902} {\bibfield  {journal} {\bibinfo
  {journal} {Journal of Chemical Physics}\ }\textbf {\bibinfo {volume} {145}},\
  \bibinfo {pages} {174112} (\bibinfo {year} {2016})}\BibitemShut {NoStop}%
\bibitem [{\citenamefont {Murakami}\ and\ \citenamefont
  {Frankcombe}(2018)}]{murakami2018}%
  \BibitemOpen
  \bibfield  {author} {\bibinfo {author} {\bibfnamefont {T.}~\bibnamefont
  {Murakami}}\ and\ \bibinfo {author} {\bibfnamefont {T.~J.}\ \bibnamefont
  {Frankcombe}},\ }\href {\doibase 10.1063/1.5046643} {\bibfield  {journal}
  {\bibinfo  {journal} {Journal of Chemical Physics}\ }\textbf {\bibinfo
  {volume} {149}},\ \bibinfo {pages} {134113} (\bibinfo {year}
  {2018})}\BibitemShut {NoStop}%
\bibitem [{\citenamefont {Green}\ \emph {et~al.}(2016)\citenamefont {Green},
  \citenamefont {Grigolo}, \citenamefont {Ronto},\ and\ \citenamefont
  {Shalashilin}}]{green2016}%
  \BibitemOpen
  \bibfield  {author} {\bibinfo {author} {\bibfnamefont {J.~A.}\ \bibnamefont
  {Green}}, \bibinfo {author} {\bibfnamefont {A.}~\bibnamefont {Grigolo}},
  \bibinfo {author} {\bibfnamefont {M.}~\bibnamefont {Ronto}}, \ and\ \bibinfo
  {author} {\bibfnamefont {D.~V.}\ \bibnamefont {Shalashilin}},\ }\href
  {\doibase http://dx.doi.org/10.1063/1.4939205} {\bibfield  {journal}
  {\bibinfo  {journal} {Journal of Chemical Physics}\ }\textbf {\bibinfo
  {volume} {144}},\ \bibinfo {eid} {024111} (\bibinfo {year}
  {2016})}\BibitemShut {NoStop}%
\bibitem [{\citenamefont {Green}\ and\ \citenamefont
  {Shalashilin}(2015)}]{green2015}%
  \BibitemOpen
  \bibfield  {author} {\bibinfo {author} {\bibfnamefont {J.~A.}\ \bibnamefont
  {Green}}\ and\ \bibinfo {author} {\bibfnamefont {D.~V.}\ \bibnamefont
  {Shalashilin}},\ }\href {\doibase
  http://dx.doi.org/10.1016/j.cplett.2015.10.073} {\bibfield  {journal}
  {\bibinfo  {journal} {Chemical Physics Letters}\ }\textbf {\bibinfo {volume}
  {641}},\ \bibinfo {pages} {173} (\bibinfo {year} {2015})}\BibitemShut
  {NoStop}%
\bibitem [{\citenamefont {Streltsov}()}]{mctdhb_lab}%
  \BibitemOpen
  \bibfield  {author} {\bibinfo {author} {\bibfnamefont {A.~I. e.~a.}\
  \bibnamefont {Streltsov}},\ }\href {http://mctdhb.org} {\enquote {\bibinfo
  {title} {The multiconfigurational time-dependent hartree for bosons
  package},}\ }\bibinfo {note} {Http://mctdhb.org}\BibitemShut {NoStop}%
\bibitem [{\citenamefont {Shalashilin}\ \emph {et~al.}(2004)\citenamefont
  {Shalashilin}, \citenamefont {Child},\ and\ \citenamefont
  {Clary}}]{shalashilin2004b}%
  \BibitemOpen
  \bibfield  {author} {\bibinfo {author} {\bibfnamefont {D.~V.}\ \bibnamefont
  {Shalashilin}}, \bibinfo {author} {\bibfnamefont {M.~S.}\ \bibnamefont
  {Child}}, \ and\ \bibinfo {author} {\bibfnamefont {D.~C.}\ \bibnamefont
  {Clary}},\ }\href {\doibase 10.1063/1.1650299} {\bibfield  {journal}
  {\bibinfo  {journal} {Journal of Chemical Physics}\ }\textbf {\bibinfo
  {volume} {120}},\ \bibinfo {pages} {5608} (\bibinfo {year}
  {2004})}\BibitemShut {NoStop}%
\bibitem [{\citenamefont {Sakmann}\ \emph {et~al.}(2014)\citenamefont
  {Sakmann}, \citenamefont {Streltsov}, \citenamefont {Alon},\ and\
  \citenamefont {Cederbaum}}]{sakmann2014}%
  \BibitemOpen
  \bibfield  {author} {\bibinfo {author} {\bibfnamefont {K.}~\bibnamefont
  {Sakmann}}, \bibinfo {author} {\bibfnamefont {A.~I.}\ \bibnamefont
  {Streltsov}}, \bibinfo {author} {\bibfnamefont {O.~E.}\ \bibnamefont {Alon}},
  \ and\ \bibinfo {author} {\bibfnamefont {L.~S.}\ \bibnamefont {Cederbaum}},\
  }\href {\doibase 10.1103/PhysRevA.89.023602} {\bibfield  {journal} {\bibinfo
  {journal} {Physical Review A}\ }\textbf {\bibinfo {volume} {89}},\ \bibinfo
  {eid} {023602} (\bibinfo {year} {2014})}\BibitemShut {NoStop}%
\bibitem [{\citenamefont {Grigolo}\ \emph {et~al.}(2016)\citenamefont
  {Grigolo}, \citenamefont {Viscondi},\ and\ \citenamefont
  {de~Aguiar}}]{grigolo2016}%
  \BibitemOpen
  \bibfield  {author} {\bibinfo {author} {\bibfnamefont {A.}~\bibnamefont
  {Grigolo}}, \bibinfo {author} {\bibfnamefont {T.~F.}\ \bibnamefont
  {Viscondi}}, \ and\ \bibinfo {author} {\bibfnamefont {M.~A.~M.}\ \bibnamefont
  {de~Aguiar}},\ }\href {\doibase 10.1063/1.4942926} {\bibfield  {journal}
  {\bibinfo  {journal} {Journal of Chemical Physics}\ }\textbf {\bibinfo
  {volume} {144}},\ \bibinfo {pages} {094106} (\bibinfo {year}
  {2016})}\BibitemShut {NoStop}%
\bibitem [{\citenamefont {Shalashilin}(2018)}]{shalashilin2018}%
  \BibitemOpen
  \bibfield  {author} {\bibinfo {author} {\bibfnamefont {D.~V.}\ \bibnamefont
  {Shalashilin}},\ }\href {\doibase 10.1063/1.5023209} {\bibfield  {journal}
  {\bibinfo  {journal} {Journal of Chemical Physics}\ }\textbf {\bibinfo
  {volume} {148}},\ \bibinfo {pages} {194109} (\bibinfo {year}
  {2018})}\BibitemShut {NoStop}%
\bibitem [{\citenamefont {Alon}\ \emph {et~al.}(2007)\citenamefont {Alon},
  \citenamefont {Streltsov},\ and\ \citenamefont {Cederbaum}}]{alon2007}%
  \BibitemOpen
  \bibfield  {author} {\bibinfo {author} {\bibfnamefont {O.~E.}\ \bibnamefont
  {Alon}}, \bibinfo {author} {\bibfnamefont {A.~I.}\ \bibnamefont {Streltsov}},
  \ and\ \bibinfo {author} {\bibfnamefont {L.~S.}\ \bibnamefont {Cederbaum}},\
  }\href {\doibase http://dx.doi.org/10.1063/1.2771159} {\bibfield  {journal}
  {\bibinfo  {journal} {Journal of Chemical Physics}\ }\textbf {\bibinfo
  {volume} {127}},\ \bibinfo {eid} {154103} (\bibinfo {year}
  {2007})}\BibitemShut {NoStop}%
\bibitem [{\citenamefont {Cao}\ \emph {et~al.}(2017)\citenamefont {Cao},
  \citenamefont {Bolsinger}, \citenamefont {Mistakidis}, \citenamefont
  {Koutentakis}, \citenamefont {Krönke}, \citenamefont {Schurer},\ and\
  \citenamefont {Schmelcher}}]{cao2017}%
  \BibitemOpen
  \bibfield  {author} {\bibinfo {author} {\bibfnamefont {L.}~\bibnamefont
  {Cao}}, \bibinfo {author} {\bibfnamefont {V.}~\bibnamefont {Bolsinger}},
  \bibinfo {author} {\bibfnamefont {S.~I.}\ \bibnamefont {Mistakidis}},
  \bibinfo {author} {\bibfnamefont {G.~M.}\ \bibnamefont {Koutentakis}},
  \bibinfo {author} {\bibfnamefont {S.}~\bibnamefont {Krönke}}, \bibinfo
  {author} {\bibfnamefont {J.~M.}\ \bibnamefont {Schurer}}, \ and\ \bibinfo
  {author} {\bibfnamefont {P.}~\bibnamefont {Schmelcher}},\ }\href {\doibase
  10.1063/1.4993512} {\bibfield  {journal} {\bibinfo  {journal} {Journal of
  Chemical Physics}\ }\textbf {\bibinfo {volume} {147}},\ \bibinfo {pages}
  {044106} (\bibinfo {year} {2017})}\BibitemShut {NoStop}%
\bibitem [{\citenamefont {Edwards}\ \emph {et~al.}(1996)\citenamefont
  {Edwards}, \citenamefont {Dodd}, \citenamefont {Clark},\ and\ \citenamefont
  {Burnett}}]{edwards1996}%
  \BibitemOpen
  \bibfield  {author} {\bibinfo {author} {\bibfnamefont {M.}~\bibnamefont
  {Edwards}}, \bibinfo {author} {\bibfnamefont {R.~J.}\ \bibnamefont {Dodd}},
  \bibinfo {author} {\bibfnamefont {C.~W.}\ \bibnamefont {Clark}}, \ and\
  \bibinfo {author} {\bibfnamefont {K.}~\bibnamefont {Burnett}},\ }\href
  {\doibase 10.6028/jres.101.055} {\bibfield  {journal} {\bibinfo  {journal}
  {Journal of research of the National Institute of Standards and Technology}\
  }\textbf {\bibinfo {volume} {101}},\ \bibinfo {pages} {553} (\bibinfo {year}
  {1996})}\BibitemShut {NoStop}%
\end{thebibliography}

%merlin.mbs apsrev4-1.bst 2010-07-25 4.21a (PWD, AO, DPC) hacked
%Control: key (0)
%Control: author (72) initials jnrlst
%Control: editor formatted (1) identically to author
%Control: production of article title (-1) disabled
%Control: page (0) single
%Control: year (1) truncated
%Control: production of eprint (0) enabled
%

\end{document}